\documentclass[12pt,letterpaper]{iopartag}
\fontfamily{cmr}
\fontseries{b}
\selectfont
\usepackage[latin1]{inputenc}
\usepackage[dvips]{graphicx}
\usepackage{amsmath}
\usepackage{iopams}
\usepackage{exscale}
\begin{document}
\eqnobysec

\jl{1}
\title{Semiclassical Propagation of Wavepackets with Complex and Real Trajectories}
\author{M. A. M. de Aguiar$^1$, M. Baranger$^{2,3}$, L. Jaubert$^4$,
Fernando Parisio$^1$ ~and A.D. Ribeiro$^1$}

\address{$^1$ Instituto de F\'{\i}sica `Gleb Wataghin',
Universidade Estadual de Campinas, 13083-970, Campinas, Brazil}
\address{$^2$ Physics Department, University of Arizona, Tucson, AZ
85719, USA}
\address{$^3$ Center for Theoretical Physics, Laboratory for Nuclear
Science and Department of Physics, Massachusetts Institute of
Technology, Cambridge, MA 02139, USA}
\address{$^4$ Laboratoire de Physique, \'{E}cole Normale
Sup\'{e}rieure de Lyon, 46 All\'{e}e d'Italie, 69364 Lyon Cedex 07,
France.}

\begin{abstract}

We consider a semiclassical approximation, first derived by Heller
and coworkers, for the time evolution of an
originally gaussian wave packet in terms of complex trajectories. We also
derive additional approximations replacing the complex trajectories by real
ones. These yield three different semiclassical formulae involving different
real trajectories. One of these formulae is Heller's thawed gaussian
approximation. The other approximations are non-gaussian and may involve
several trajectories determined by mixed initial-final conditions. These
different formulae are tested for the cases of scattering by a hard wall,
scattering by an attractive gaussian potential, and bound motion in a quartic
oscillator. The formula with complex trajectories gives good results in all
cases. The non-gaussian approximations with real trajectories work well in
some cases, whereas the thawed gaussian works only in very simple situations.

\end{abstract}

\section{Introduction}

The Feynman propagator $\langle x_f|K(T)|x_i\rangle = \langle
x_f|\exp{(-iHT/\hbar)}|x_i\rangle$ can be interpreted as the time evolution of
a wave function $\psi(x_f,T)$ that is initially localized at $x_i$,
$\psi(x,0)=\delta(x-x_i)$. In the semiclassical limit $\langle
x_f|K(T)|x_i\rangle$ can be obtained from the classical trajectories
connecting the initial point $x_i$ to the final point $x_f$ in time $T$. The
formula which does this is known as the Van Vleck approximation.

One is more likely, however, to be interested in the propagation of a smooth
wavepacket, say $\psi_0(x)$, rather than an eigenfunction of the position or
momentum operators. The straightforward way to accomplish this propagation,
given the knowledge of $K(T)$, is with an extra integration, thus
\begin{align}
   \label{ba1}
       \psi(x_f,T) = \int \langle x_f|K(T)|x_i\rangle \,
           {\rm d}x_i\langle x_i|\psi_0\rangle \;.
\end{align}
In the semiclassical limit this formula becomes
\begin{align}
   \label{ba1x}
     \psi_{\rm sc}(x_f,T) = \int\langle x_f|K(T)|x_i\rangle_{\rm Van Vleck} \:
           {\rm d} x_i \langle x_i| \psi_0 \rangle \;.
\end{align}

One problem with this approach is that there are usually several classical
trajectories going from $x_i$ to $x_f$ and they are hard to find, the
so-called root-search difficulty.  We shall return to a discussion of this
point in section 7. In the present paper, we shall accept the root-search
problem and start from equation (\ref{ba1x}).

The main purpose of this paper is to derive various approximations for
Eq.(\ref{ba1x}) and to test them in simple problems. Hence we
restrict ourselves here to systems with one degree of freedom. In
principle, all the expressions that we obtain are easily
generalized to multi-dimensional systems. In practice, of course,
the difficulty of the numerical calculations increases very fast
with the dimensionality, and it also depends on the actual
approximation used.

We shall evaluate the integral by the method of steepest descent
which is also often called, somewhat inaccurately, stationary
phase. We shall see that the stationary point is generally
complex, giving rise to complex classical dynamics.  Such a
calculation was performed first by Heller and collaborators
\cite{Hel87,Hel88}.  The result that they derived can be shown to
be identical to our Eq.(\ref{ba14}).  They used it to calculate
the motion of a wavepacket in the Morse potential.  Their work
does not seem to have received from physicists all the recognition
that it deserves.  Perhaps this was because there were two rather
lengthy papers which contained many somewhat unfamiliar notions.
Their first paper \cite{Hel87} reached the result through
heuristic arguments and the second one \cite{Hel88}, though more
direct, made many references to the first.  In our paper, on the
other hand, we shall present in section 2 a very simple and very
short derivation of Heller's formula, Eq.(\ref{ba14}).

This basic complex approximation will be the starting point for all other
developments of this paper.  We shall show that it can be handled for
relatively simple problems, but we shall also derive three subsequent
approximations involving only real trajectories, which are different for each
of the three cases.  One case yields the well-known Heller Thawed Gaussian
Approximation (TGA) \cite{Hell75}.  The other two give different results and
we shall show that they can be quite accurate in some situations.  We shall
illustrate the application of the various semiclassical formulae with two very
simple examples (section 4) and two not so simple ones (sections 5 and 6).  We
hope that our derivation of the basic formula and our examples comparing the
several approximations will stimulate new interest in semiclassical formulae
derived from complex trajectories.

In order to simplify our calculations we take for initial state
$|\psi_0\rangle$ a gaussian wavefunction with average position and momentum
$q$ and $p$ respectively, and position uncertainty $\Delta q=b/\sqrt{2}$. This
state is also the coherent state $|z\rangle$ of a harmonic oscillator of
mass $\mu$ and frequency $\omega$. It is defined by
\begin{equation}
   \label{glg48}
   |z\rangle = e^{-\frac{1}{2}|z|^2}\; e^{z\hat{a}^\dagger}|0\rangle
\end{equation}
with $|0\rangle$ the harmonic oscillator ground state and
\begin{equation}
   \label{glg49}
   \hat{a}^\dagger = \frac{1}{\sqrt{2}}\left( \frac{\hat{q}}{b}-i
               \,\frac{\hat{p}}{c} \right), \qquad
   z =  \frac{1}{\sqrt{2}}\left( \frac{q}{b}+i
               \,\frac{p}{c} \right).
\end{equation}
In the above $\hat q$, $\hat p$, and $\hat{a}^\dagger$ are operators;
$q$ and $p$ are real numbers; $z$ is complex.  The parameters
\begin{equation}
   \label{glg47}
   b = {(\hbar/ \mu \omega )}^{\frac{1}{2}}\qquad \mbox{and} \qquad
   c = {(\hbar \mu \omega )}^{\frac{1}{2}}
\end{equation}
define the length and momentum scales, respectively, and their product is
$\hbar$. The final form of Eq.(\ref{ba1x}), our semiclassical approximation
for wavepacket propagation, is then
\begin{align}
   \label{ba3a}
       \langle x_f|K(T)|z\rangle
    \approx  \int\langle x_f|K(T)|x_i\rangle_{{\rm Van Vleck}} \,
       {\rm d}x_i\langle x_i|z\rangle  \equiv \psi_{\rm sc}(x_f,z;T) \; .
\end{align}

\section{Approximation with complex trajectories}

Before we perform the integral in equation (\ref{ba3a}) by
stationary phase, we recall the relations between the elements of
the tangent matrix $m$ and the second derivatives of the action
function $S(x_f,x_i;T)$ computed along the real classical trajectory
connecting $x_i$ to $x_f$ in time $T$.  Given a classical trajectory $X(t)$,
$P(t)$ with $X(0)=x_i$ and $X(T)=x_f$, its tangent matrix $m$ connects, in the
linearized approximation, a small initial displacement $\delta x_i,\delta p_i$
about the trajectory at $t=0$ to the propagated displacements $\delta
x_f,\delta p_f$ at time $T$. The relation between $m$ and the second
derivatives of the action is
\begin{align}
   \label{ba4aa}
        \begin{pmatrix}
           \frac{\delta x_f}{b} \\ \\
           \frac{\delta p_f}{c}
        \end{pmatrix}
    =   \begin{pmatrix}
          -\frac{S_{ii}}{S_{if}} & -\frac{c}{b} \frac{1}{S_{if}} \\ \\
           \frac{b}{c} \left(S_{if} -S_{ff}\frac{S_{ii}}{S_{if}}\right)
             & -\frac{S_{ff}}{S_{if}}
           \end{pmatrix}
         \begin{pmatrix}
           \frac{\delta x_i}{b} \\ \\
           \frac{\delta p_i}{c}
         \end{pmatrix}
    \equiv  \begin{pmatrix}
              m_{qq} & m_{qp} \\ \\
              m_{pq}  & m_{pp}
            \end{pmatrix}
            \begin{pmatrix}
              \frac{\delta x_i}{b} \\ \\
              \frac{\delta p_i}{c}
            \end{pmatrix}
\end{align}
where $S_{ii} \equiv \partial^2 S/\partial {x_i}^2$, $S_{if} = S_{fi}
\equiv \partial^2 S/\partial {x_i} \partial{x_f}$ and $S_{ff} \equiv
\partial^2 S/\partial {x_f}^2$. A derivation of this formula can be found,
for instance, in sec.(2.6) of \cite{Bar01}. Notice that we define
$m$ using the coherent state scales $b$ and $c$. Inverting this
equation we obtain
\begin{align}
   \label{ba4a}
S_{ii} = \frac{c}{b} \, \frac{m_{qq}}{m_{qp}}, \qquad S_{if} =
-\frac{c}{b} \, \frac{1}{m_{qp}}, \qquad S_{ff} = \frac{c}{b} \,
\frac{m_{pp}}{m_{qp}} \;.
\end{align}

We shall write our final results in terms of elements of the
tangent matrix $m$, rather than in terms of derivatives of the
action. In this notation the Van Vleck propagator is
\begin{align}
   \label{ba5}
\langle x_f|K(T)|x_i\rangle_{{\rm Van Vleck}}
   &= \frac{1}{b\sqrt{2\pi m_{qp}}} \;  e^{i S /\hbar - i\pi/4}\; .
\end{align}
For short times $m_{qp}$ is positive and the square root is well defined. For
longer times $m_{qp}$ may become negative by going through zero. This happens
at the focal points. At these points the Van Vleck formula diverges and must
be replaced by a higher order approximation involving Airy functions.
Sufficiently away from the foci the approximation becomes good again, as long
as one replaces $m_{qp}$ by its modulus and subtracts a phase $\pi/2$ for
every focus encountered along the trajectory. We shall not write these
so-called Morse phases explicitly because they will cancel out in our final
formula.

The overlap $\langle x_i|z\rangle$ is given by
\begin{align}
   \label{ba6}
\langle x_i|z\rangle &= \pi^{-\frac{1}{4}} b^{-\frac{1}{2}}
                       \exp\left(-\frac{{(x_i-q)}^2}{2 b^2}\right)
                       \exp\left(\frac{i}{\hbar}p(x_i - q/2)\right) \;.
\end{align}
In the semiclassical limit the propagated wavepacket (\ref{ba3a})
is then
\begin{align}
   \label{ba7}
        \psi_{\rm sc}(x_f,z;T)
    =  \int  \frac{\exp{\{\Phi\}}}{b^{3/2} \pi^{1/4} \sqrt{2\pi m_{qp}}} \;
     {\rm d} x_i
\end{align}
with
\begin{align}
   \label{ba8}
     \Phi(x_f,x_i,T) = \frac{i}{\hbar} \left[ S + p(x_i-q/2)\right]
     - \frac{(x_i-q)^2}{2 b^2} -i\pi/4 \;.
\end{align}

We shall now approximate the integration over $x_i$ by the stationary phase
method. To do so we assume that the pre-factor in (\ref{ba7}) is a
slowly-varying function of $x_i$, which means that the stationary point $x_0$
will be computed by imposing zero variation on $\Phi$ alone. The pre-factor
will be simply computed at $x_0$ and will not be expanded. We refer to
section 3.3 of \cite{Bar01} for a discussion of this procedure. We need to
calculate the first and second derivatives of $\Phi$. The first derivative is
\begin{align}
   \label{ba9}
     \Phi' \equiv \frac{\partial \Phi}{\partial x_i} =
     \frac{i}{\hbar} (p-p_i) - \frac{(x_i-q)}{b^2}
\end{align}
where we have defined
\begin{align}
   \label{ba11}
     p_i(x_f,x_i\,;T) =  - \frac{\partial S}{\partial x_i} \;.
\end{align}
The second derivative is
\begin{align}
   \label{ba10}
     \Phi'' \equiv \frac{\partial^2 \Phi}{\partial {x_i}^2} =
     - \frac{i}{\hbar} \frac{ \partial p_i}{\partial x_i}
      - \frac{1}{b^2} \;.
\end{align}
The stationary position $x_0(x_f,T)$, solution of the stationary
phase condition $\Phi'=0$, is given by the equation
\begin{align}
   \label{ba9x}
     \frac{x_0-q}{b} +i \frac{p_0-p}{c} = 0
\end{align}
where we have used $c=\hbar/b$ and we have made the definition
\begin{align}
   \label{ba11x}
     p_0(x_f\,;T) = p_i(x_f,x_0\,;T) \;.
\end{align}
Equation (\ref{ba9x}) makes it clear that $x_0$ is usually
complex, and therefore the stationary classical trajectory itself
is complex. It leaves $x_0$ at time $0$ with momentum $p_0$,
also complex, and it arrives at $x_f$, which is real, at time $T$. It goes
without saying that, for such a complex trajectory, the tangent matrix $m$
is complex as well.

To integrate over $x_i$, we expand $\Phi$ around $x_0$ to second
order, which means that we write $\Phi \approx \Phi(x_0) +
\Phi''(x_0) \, (x_i-x_0)^2 \,/2\;$. If Im$(\partial
p_i/\partial x_i)$ is less than $\hbar/b^2$ at $x_0$, then
Re$(\Phi)$ has a negative definite quadratic term and the integral
can be performed in the gaussian approximation. Assuming this to
be the case we find
\begin{align}
\label{ba12} \psi_{\rm sc}(x_f,z;T) = \frac{1}{b^{3/2} \pi^{1/4}
\sqrt{2\pi m_{qp}}} \; \sqrt{\frac{-2\pi}{\Phi''(x_0)}} \;
\exp{\{\Phi(x_0)\}} \;.
\end{align}
Finally, using equations (\ref{ba10}), (\ref{ba11}), (\ref{ba4a}),
we notice that
\begin{align}
   \label{ba13}
\Phi''(x_0) = \frac{i}{\hbar} S_{ii} -\frac{1}{b^2}
              = \frac{i}{b^2} \frac{m_{qq}+i m_{qp}}{m_{qp}} \;.
\end{align}
This simplifies the pre-factor in (\ref{ba12}) and we get
\begin{multline}
   \label{ba14}
      \psi_{\rm CT}(x_f,z\,;T)
    =  \frac{1}{b^{1/2} \pi^{1/4}} \frac{1}{\sqrt{m_{qq}+im_{qp}}}
     \\
    \times \exp{ \left[ \frac{i}{\hbar} S(x_f,x_0\,;T) +
          \frac{i}{\hbar} p(x_0-q/2) - \frac{(x_0-q)^2}{2b^2} \right]} \;.
\end{multline}
We have changed the subscript on $\psi$ from ``sc'' to ``CT'' to indicate that
this semiclassical approximation is calculated with complex trajectories.

Equation (\ref{ba14}), the semiclassical limit of the propagated wavepacket,
is the Heller result \cite{Hel87,Hel88}. It is the basic result from which all
other approximations in this paper are derived.  It is {\em not} an initial
value representation and it may involve more than one complex classical
trajectory. At first glance one might think that this formula represents a
frozen gaussian, since the quadratic term in the exponent has a fixed width
$b$. A closer look, however, reveals that, since the classical trajectory is
complex, all other terms in the exponent also contribute a real part, which
changes the effective width: the complex character of the stationary
trajectory makes the wavepacket thaw and, at the same time, assume
non-gaussian shapes.  Notice that no Morse index or phases of $\pi/4$ appear
in (\ref{ba14}). At $T=0$ one has $m_{qq}=1$, $m_{qp}=0\,:$ one chooses the
positive square root and the overlap (\ref{ba6}) is recovered.  For other
times one simply follows the phase of the complex number $m_{qq}+im_{qp}$ to
get the phase of the pre-factor.

Before we close this section we re-write the boundary conditions
for the complex trajectories in a more convenient form. The
stationary trajectory starts at $x_0$ (usually complex) and ends
at $x_f$ (always real). The initial momentum (also complex) is $p_0$,
given by (\ref{ba11x}) and (\ref{ba11}). Since the coordinates $X(t)$ and
$P(t)$ along the trajectory are both going to be complex, it is convenient to
define the variables
\begin{eqnarray}
    u = \frac{1}{\sqrt{2}}
    \left( \frac{X}{b} + i \frac{P}{c} \right) \qquad \qquad
    v = \frac{1}{\sqrt{2}}
    \left( \frac{X}{b} - i \frac{P}{c} \right)
    \label{uv}
\end{eqnarray}
where $u \neq v^{*}$ in general. In this notation Hamilton's equations read
\begin{equation}
i \hbar \dot{u} =  \frac{\partial H}{\partial v}
  \qquad  \qquad \mathrm{and} \qquad \qquad
- i \hbar \dot{v} =  \frac{\partial H}{\partial u} \; .
\label{Hamiluv}
\end{equation}
Finally the stationary condition (\ref{ba9x}) can be re-written as
\begin{align}
   \label{bc}
     \frac{x_0}{b} + i \frac{p_0}{c} = \frac{q}{b} + i\frac{p}{c}
\end{align}
which is equivalent to $u(0)=z$. Therefore the stationary
trajectories are solutions of Hamilton's equations satisfying
\begin{equation}
     u(0) = z \qquad \qquad \mbox{and} \qquad \qquad  X(T) = x_f \; .
     \label{bcf}
\end{equation}


\section{Approximations with real trajectories}

The stationary phase approximation of the previous section replaces the
integral over a continuum of classical real trajectories by a few complex
ones. Finding complex trajectories, however, is generally harder than finding
real trajectories and one is tempted to look for further approximations to
Eq.(\ref{ba14}) in terms of real trajectories only. These approximations
should be good if the stationary complex trajectory is sufficiently close to a
real trajectory.

What real trajectory? According to (\ref{bcf}), the complex trajectory is
determined by two pieces of data: a final position $x_f$ which is real and an
initial coordinate $u(0) = z$ which is complex; see (\ref{bc}). The latter is
made up of a real position $q$ and a real momentum $p$. Thus we have a total
of three real parameters, $x_f$, $q\,,$ and $p\,,$ that could be used to
determine a real trajectory. Since it actually takes only two parameters, we
have three obvious ways of choosing a real trajectory that might sometimes be
a good approximation to the complex one. The first way is to give the two
initial conditions $(q,p)$, which are the center of the initial wavepacket in
phase space. The other two ways use the final position $x_f$ with a single one
of the initial conditions, either $(x_f,q)$ or $(x_f,p)$. Each of these three
choices leads to a possible approximation in terms of a real trajectory and,
as we shall see, they are all different. In this section we shall derive
formulae for each of these three cases. The extent of their validity will be
examined in later sections.

We can try to get a rough sense of how close these three real trajectories are
to the unique complex one. The latter goes through $x_f:$ hence the
$(x_f,q)$ and $(x_f,p)$ trajectories have at least the merit of being at the
right position for the final time $T$. To judge the closeness at time 0, we
can look at formula (\ref{ba14}) which tells us that, if Im$(x_0)$ is
sufficiently small compared to $b$, the semiclassical wave function at time
$T$ will be very small unless Re$(x_0)-q$ is of order $b$ or smaller. This is
a small distance in the semiclassical limit since $b$ is of order
$\hbar^{1/2}$. We can also estimate the closeness of the initial momenta by
looking at the stationary phase requirement (\ref{ba9x}), separating its real
and imaginary parts thus
\begin{align}
\begin{split}
\frac{{\rm Re}\,x_0-q}{b} &= \frac{{\rm Im}\,p_0}{c}   \\
\frac{{\rm Re}\,p_0-p}{c} &= -\frac{{\rm Im}\,x_0}{b} \; .
\end{split}
\label{reim}
\end{align}
Again assuming Im$(x_0)$ sufficiently small compared to $b$, we see from the
second equation that Re$(p_0)-p$ is of order $c$ or smaller, and $c$ is again
proportional to $\hbar^{1/2}$. The conclusion is that a semiclassical
approximation in terms of real classical trajectories might be useful if
conditions are right.

\subsection{The central trajectory approximation: Heller's TGA}

To begin, we choose the real trajectory defined by the two initial conditions
$(q,p)$, the central trajectory of the packet. We call $(q_T, \, p_T)$ the
real end point of this trajectory at time $T$. This trajectory will now be the
0-order approximation. All values of $S$ and its derivatives that occur in the
equations, including the $m$ matrix, will be taken for this trajectory, unless
we state otherwise explicitly. All these quantities are now real.

We write
\begin{align}
   \label{ba15a}
     x_0 & = q + \Delta x_i \\
   \label{ba15b}
     x_f & = q_T + \Delta x_f \\
   \label{ba15c}
     p_0 & \equiv -\frac{\partial S}{\partial x_i}(x_f,x_0\,;T) = p + \Delta p_i
       = p - S_{ii} \Delta x_i - S_{if} \Delta x_f \\
   \label{ba15d}
     p_f & \equiv +\frac{\partial S}{\partial x_f}(x_f,x_0\,;T) = p_T+\Delta p_f
       = p_T + S_{fi}\Delta x_i + S_{ff} \Delta x_f \;.
\end{align}
The stationary phase condition (\ref{ba9x}), can be re-written as
$\Delta x_i/b + i \Delta p_i/c  = 0$. Then equation (\ref{ba15c})
can be solved for $\Delta x_i$
\begin{align}
  \label{ba16}
    \Delta x_i =  -\frac{S_{if}}{S_{ii}+ic/b} \, \Delta x_f
              = \frac{1}{m_{qq}+i m_{qp}} \, \Delta x_f
\end{align}
To simplify the notation, the complex number $m_{qq}+i m_{qp}$ will be called
$m_+\;$.

We proceed to expand the exponent in (\ref{ba14}) about the real
trajectory. As before, the pre-factor is assumed to be a
slowly-varying function and will be simply computed at the real
trajectory. We write
\begin{align}
  \label{ba17}
    & \frac{i}{\hbar} S(x_f,x_0;T) +
      \frac{i}{\hbar}p(x_0-q/2) - \frac{(x_0-q)^2}{2b^2} \nonumber \\
    & \approx \frac{i}{\hbar}S
      -  \frac{i}{\hbar}p\Delta x_i + \frac{i}{\hbar}p_T\Delta x_f
       \nonumber \\
    & \quad +\frac{i}{2\hbar}(S_{ii} \Delta {x_i}^2 + 2S_{if} \Delta x_i
      \Delta x_f + S_{ff} \Delta {x_f}^2)  \nonumber \\
    & \quad + \frac{i}{\hbar}pq/2 +\frac{i}{\hbar}p\Delta x_i
      - \frac{\Delta {x_i}^2}{2b^2} \;.
\end{align}
The linear terms in $\Delta x_i$ cancel. The quadratic terms can
all be written in terms of $\Delta x_f = (x_f-q_T)$ using (\ref{ba16}).
All second derivatives of the action can be written in terms of the
tangent matrix using (\ref{ba4a}). Collecting all quadratic terms we get:
\begin{align}
  \label{ba17a}
 & \frac{i}{2bc}\left(S_{ii} \Delta {x_i}^2 + 2S_{if} \Delta x_i
      \Delta x_f + S_{ff} \Delta {x_f}^2\right) - \frac{\Delta {x_i}^2}{2b^2}
      \nonumber \\
 =& \frac{\Delta {x_f}^2}{2b^2} \,
     \left[i \, \frac{m_{qq}}{m_{qp}} \, \frac{1}{{m_+}^2}
    - 2i \, \frac{1}{m_{qp}} \, \frac{1}{m_+}
    + i \, \frac{m_{pp}}{m_{qp}}
    - \frac{1}{{m_+}^2}
     \right] \nonumber \\
 =& \frac{1}{{m_+}^2 m_{qp}} \, \frac{\Delta {x_f}^2}{2b^2} \left[
    i m_{qq} -2i(m_{qq}+i m_{qp}) + i m_{pp} {m_+}^2 - m_{qp} \right]
    \nonumber \\
 =& \frac{1}{{m_+}^2 m_{qp}} \, \frac{\Delta {x_f}^2}{2b^2} \left[
    -i m_{qq} + m_{qp} + i m_{pp} {m_+}^2 \right] \nonumber \\
 =& -\frac{i}{{m_+} m_{qp}} \, \frac{\Delta {x_f}^2}{2b^2} \left[
     1-{m_+} m_{pp} \right] = -\frac{\Delta {x_f}^2}{2 b^2} \,
     \frac{m_{pp}-im_{pq}}{m_{qq}+im_{qp}}
\end{align}
where in the last equality we have used
$m_{qq}m_{pp}-m_{qp}m_{pq}=1$ .

With these simplifications the semiclassical wavepacket formula becomes
\begin{multline}
  \label{ba18}
     \psi_{qp}(x_f,z\,;T) =
      \frac{b^{-1/2} \pi^{-1/4} } {\sqrt{m_{qq}+im_{qp} } }
      \exp{ \left[ \frac{i}{\hbar} S(q_T,q\,;T) + \frac{i pq}{2\hbar}
      \right.} \\
      \left. +\frac{i}{\hbar} p_T(x_f-q_T) - \frac{1}{2}
      \left( \frac{m_{pp}-im_{pq} }{m_{qq}+im_{qp} } \right) \,
      \left( \frac{x_f-q_T}{b} \right)^2 \right] \;.
\end{multline}
The subscript $qp$ on $\psi$ indicates the variables used to compute the real
trajectory. The spreading of the propagated wavepacket is now explicit in the
coefficient of the gaussian term. This is exactly Heller's thawed gaussian
approximation or TGA \cite{Hell75}. As discussed at the end of section 7, a
similar approximation \cite{Bar01} can be obtained with coherent state path
integrals.

\subsection{The approximation by trajectory $q \longrightarrow x_f$}

We consider now as 0-order the real trajectory which starts at $q$ and
ends at $x_f$ after time $T$. We call $p_i$ its initial momentum, which is a
function of $x_f$, $q$, and $T$, and we write
\begin{align}
\begin{split}
x_0&=q+\Delta x_i\\
p_0& \equiv -\frac{\partial S}{\partial x_i}(x_f,x_0\,;T)
=p_i+\Delta p_i=p_i-S_{ii}\Delta x_i \; .
\end{split}
\label{deltap1}
\end{align}
Notice that the complete expansion of $p_0$ to first order
should be $p_i-S_{ii}\Delta x_i-S_{if}\Delta x_f$ but, as $x_f$ is
fixed, $\Delta x_f = 0$. Eq.(\ref{ba9x}) gives us the relation between
$\Delta x_i$ and $\Delta p_i$
\begin{eqnarray}
i\frac{(x_0-q)}{b}=\frac{p_0-p}{c}=
\frac{p_0-p_i}{c}+\frac{p_i-p}{c}
 \label{deltap2}
\end{eqnarray}
Thanks to Eqs.(\ref{deltap1}) and (\ref{deltap2}) we obtain
\begin{eqnarray}
\Delta p_i=\frac{ic}{b}\Delta x_i-(p_i-p)=-S_{ii}\Delta x_i
\end{eqnarray}
which gives, in terms of the tangent matrix,
\begin{eqnarray}
\Delta x_i=\frac{(p_i-p)}{S_{ii}+\dfrac{ic}{b}}= \frac{b}{c}
\; \frac{m_{qp}}{m_{qq}+im_{qp}}(p_i-p) \; . \label{deltax}
\end{eqnarray}
We expand the exponent of Eq.(\ref{ba14}) around this real trajectory. Once
again we take the action S, its derivatives $S_i$, $S_f$, and its second
derivatives $S_{ii}$, $S_{if}$, $S_{ff}$, including the $m$ matrix, for the
0-order trajectory, unless we state otherwise explicitly. We obtain
\begin{equation}
\begin{array}{ll}
 &\dfrac{i}{\hbar} S(x_f,x_0\,;T)  +\dfrac{i}{\hbar}p(x_0-q/2)
 -\dfrac{(x_0-q)^2}{2b^2}\\
&= \dfrac{i}{\hbar} \left(S
+S_i\Delta x_i +\dfrac{1}{2}S_{ii}\Delta {x_i}^{\,2} \right)
+\dfrac{i}{\hbar}p\Delta x_i +\dfrac{i}{2\hbar}pq
- \dfrac{\Delta {x_i}^{\,2}}{2b^2}\\
&= \dfrac{i}{\hbar} S + \dfrac{i}{2\hbar}pq
+\dfrac{i}{\hbar}(p-p_i)\Delta x_i +\dfrac{1}{2}
\left(\dfrac{iS_{ii}}{\hbar}-\dfrac{1}{b^2}\right)\Delta {x_i}^{\,2}\\
&= \dfrac{i}{\hbar} S +\dfrac{i}{2\hbar}pq
-\dfrac{im_{qp}}{m_{qq}+im_{qp}}\dfrac{(p_i-p)^2}{c^2} +\frac{1}{2}
\dfrac{im_{qp}}{m_{qq}+im_{qp}}\dfrac{(p_i-p)^2}{c^2} \; .
\end{array}
\end{equation}

Finally, the semiclassical propagator as a function of $q,\,p,\,x_f$, and
$p_i$, becomes
\begin{multline}
\psi_{{x\!_f}q}(x_f,z\,;T)=\frac{b^{-1/2}\pi^{-1/4}}{\sqrt{m_{qq}+im_{qp}}}
\exp\left[\frac{i}{\hbar}S(x_f,q\,;T)+\frac{i}{2\hbar}pq \right. \\
\left. -\frac{1}{2}\,\frac{im_{qp}}{m_{qq}+im_{qp}}
\left(\frac{p-p_i}{c}\right)^2\right] \; . \label{wv2}
\end{multline}
The gaussian in the exponent is now in the difference between $p_i$, the
initial momentum of the real trajectory, and $p$, the initial momentum of the
center of the wave packet. Notice that there might be more than one trajectory
satisfying the boundary conditions $X(0)=q$, $X(T)=x_f\;.$ Notice also that the
wave function is not restricted to a gaussian anymore. And that this is
{\it not} an initial value formula.

\subsection{The approximation by trajectory $p \longrightarrow x_f$}

In the same way, if we choose as 0-order the trajectory specified by $p$ and
$x_f$ instead of $q$ and $x_f$, we call $q_i$ the initial position for this
trajectory, which is a function of $x_f$, $p$, and $T$, and we write
\begin{align}
\begin{split}
x_0&=\;q_i+\Delta x_i\\
p_0&\equiv -\frac{\partial S}{\partial x_i}(x_f,x_0\,;T)
=p+\Delta p_i=p-S_{ii}\Delta x_i \; .
\end{split}
\label{defS3}
\end{align}
Thanks to eqs.(\ref{ba9x}) and (\ref{defS3}) the new expression
for $\Delta x_i$ is
\begin{eqnarray}
\Delta x_i=\frac{i m_{qp}}{m_{qq}+im_{qp}}(q-q_i) \; .
\end{eqnarray}
In the end we obtain a propagating semiclassical wavepacket different from the
other two, namely
\begin{multline}
\psi_{{x\!_f}p}(x_f,z\,;T)=\frac{b^{-1/2}\pi^{-1/4}}{\sqrt{m_{qq}+im_{qp}}}
\exp \left[\frac{i}{\hbar}S(x_f,q_i\,;T)+\frac{i}{2\hbar}pq \right. \\
\left. +\frac{i}{\hbar}p(q_i-q)
-\frac{1}{2}\,\frac{m_{qq}}{m_{qq}+im_{qp}}
\left(\frac{q-q_i}{b}\right)^2 \right] \; . \label{wv3}
\end{multline}
The gaussian in the exponent is in the difference between $q_i$, the initial
position of the real trajectory, and $q$, the initial position of the center
of the wavepacket. This formula shares many of the features of the previous
one, including not being a gaussian and not being an initial value formula.

The three real trajectory approximations we derived here are formally
similar. Among the three real variables $q$, $p$, and $x_f$, two are chosen to
fix the trajectory. Then the corresponding formula carries a gaussian damping
factor in the third variable.  In the next three sections we discuss some
examples that will help us compare these approximations with $\psi_{CT}$ which
involves complex trajectories. As usual all these formulae are exact for the
free particle and the harmonic oscillator.

\section{First applications: the free particle and the hard wall}

The exact result for the propagation of a free particle wavepacket is of
course well-known, but it can serve as a test of the formulae of secs. 2 and
3. And it is instructive to be able to see explicitly the different
trajectories involved in each approximation. The wave function for the packet
at time 0 is given by (\ref{ba6}). The hamiltonian is simply $H=P^2/2\mu$
where $\mu$ is the mass. The action is
\begin{align}
\label{fpa}
S(x_f,x_i;T) = \frac{\mu}{2} \frac{(x_f-x_i)^2}{T} \; .
\end{align}
All trajectories have constant momentum or velocity. In addition to the other
parameters we shall use $\omega=c/\mu b=\hbar/\mu b^2$, which is the frequency
of the oscillator upon which the coherent states are built, and $v=p/\mu$, the
central velocity of the wavepacket. The elements of the tangent matrix are
$m_{qq}=1, m_{qp}=\omega T, m_{pq}=0, m_{pp}=1$.

The exact expression for the propagated wavepacket follows from elementary
quantum mechanics and can be written
\begin{multline}
\psi_{\rm exact}(x_f,z;T)=
 \frac{1}{b^{1/2} \pi^{1/4}} \frac{1}{\sqrt{1+i\omega T}} \\
 \times \exp\left[-\frac{(x_f-q-vT)^2}{2b^2(1+\omega^2T^2)}(1-i\omega T)
 +\frac{i}{\hbar}p\left(x_f-\frac{q}{2}-\frac{vT}{2}\right)\right].
 \label{freeWP}
\end{multline}
It describes a gaussian packet of constant momentum $p$, whose real width
$b\sqrt{1+\omega^2T^2}$ increases with time, but whose total width is actually
complex and given by $b\sqrt{1+i\omega T}\;.$

For the approximation of sec.2, the stationary point $x_0$ of the integration,
which is the origin of the complex trajectory, and its associated momentum
$p_0$, which is the constant momentum of the trajectory, turn out to be
\begin{align}
   \label{TD1}
x_0(x_f,z;T)=\frac{x_f+i\omega T\left(q+i\,\dfrac{b}{c}\,p\right)}{1+i\omega T}
\qquad \qquad
p_0(x_f,z;T)=\frac{p+i\,\dfrac{c}{b}\,(x_f-q)}{1+i\omega T} \; .
\end{align}
One can check the two boundary conditions (\ref{bcf})
\begin{align}
   \label{2BCs}
\frac{x_0}{b}+i\,\frac{p_0}{c}=\frac{q}{b}+i\,\frac{p}{c}
\qquad\qquad\qquad\qquad    x_0 + \frac{p_0}{\mu} T = x_f \; .
\end{align}
One can also verify that, when $x_f$ has the value $q+vT$, one finds the
simple answers $x_0=q$ and $p_0=p$ : the trajectory is just the central
trajectory of the packet.

Let us look also at the three approximations with real trajectories. For the
$(q,p)$ case, formula (\ref{ba18}) contains both $q_T$ and $p_T\;.$ The former
is clearly $q_T=q+vT$, while $p_T=p$, the constant momentum. For $(x_f,q)$,
formula (\ref{wv2}), we need $p_i$ which is $p_i=\mu(x_f-q)/T$. We don't need
$p_f$, but it is obviously the same as $p_i$. Finally, for $(x_f,p)$, formula
(\ref{wv3}), we need $q_i=x_f-pT/\mu\,,$ since the constant momentum is $p\,.$
Once again, for the free particle, all four semiclassical approximations give
the exact answer.

To make things less trivial, we shall now have the wavepacket bounce
elastically against a hard wall placed at the origin. The packet arrives from
the right with a negative momentum and we assume that it starts far enough to
have no appreciable value at the wall at time 0. The only relevant spatial
region is $x_f>0\,;$ the other side of the wall does not exist for this
problem. Of course the wall could be an approximation to a very steep, but
not totally hard, potential, in which case there would be some barrier
penetration to the other side. And it would be interesting to see what happens
in the limiting process, but we shall not do that here.

The exact solution is quite simple, once one knows the exact free particle
formula (\ref{freeWP}): you take the $x_f<0$ part of the latter, you reflect it
with respect to the wall, and you give it an additional minus sign. Thus the
complete solution has two parts, both restricted to $x_f>0\,,$ the original free
particle part, and the reflected part which uses the $x_f<0$ piece of the free
particle :
\begin{align}
\label{wall}
\begin{split}
\psi_{\rm wall,\,exact}(x_f,z;T) = &\;\psi_{\rm \,free,\,exact}(x_f,z;T)
    - \psi_{\rm \,free,\,exact}(-x_f,z;T) \\
    &x_f > 0 ~ {\rm only}\,,\quad T>0
\end{split}
\end{align}
This construction ensures that the Schroedinger equation is satisfied
everywhere and that the total wave function vanishes at the wall.

Next we look at the semiclassical approximation with complex trajectories,
sec. 2. Given $x_i$ and $x_f$, both $>0$, there are two ways to go from one to
the other in time $T$, the direct way and the way which bounces off the
wall. The first trajectory is identical in all respects to a free particle's:
it has the same action (\ref{fpa}) and the process of finding the stationary
point $x_0$ of the integration over $x_i$ is the same. Hence, for this direct
trajectory, formula (\ref{ba14}) yields the free particle result, restricted
to $x_f>0\,.$ The reflected trajectory, on the other hand, has a different
action. The distance between $x_i$ and $x_f$, including reflection, is
$D=x_f+x_i$, the speed is $(x_f+x_i)/T$, the momentum $\mu(x_f+x_i)/T$, and
the energy $\mu(x_f+x_i)^2/2T^2$. Hence the action is
\begin{align}
\label{refa}
S(x_f,x_i;T) = PD-ET = \mu\dfrac{x_f+x_i}{T}(x_f+x_i)
   -\dfrac{\mu}{2}\dfrac{(x_f+x_i)^2}{T^2}T
   = \dfrac{\mu}{2}\dfrac{(x_f+x_i)^2}{T} \; .
\end{align}
It differs from (\ref{fpa}) by a sign change for one of the two coordinates.
The initial and final momenta follow from this action in the usual way:
\begin{align}
\label{moms}
p_f=\frac{\partial S}{\partial x_f}=\mu\frac{x_f+x_i}{T}
\qquad \qquad    p_i=-\frac{\partial S}{\partial x_i}=-p_f \; .
\end{align}
Compared to the direct trajectory, the only change that needs to be made when
we look for the stationary point of the $x_i$ integration and we apply formula
(\ref{ba14}), is that the constant quantity $x_f$ must be replaced everywhere
by $-x_f$. Thus we get again the free particle result, but for the value
$-x_f$ of the position. This approximation, therefore, yields the exact answer
(\ref{wall}), except for the minus sign in front of the second term. This
minus sign due to reflection is expected on very general grounds. It is a
special case of the Morse phase mentioned in sec. 2, and it is understood
easily if one looks at a soft barrier, such as a finite square step, and one
goes to the infinite limit.

Now let us try to apply to the hard wall the approximation with the $(q,p)$
real trajectory of sec. 3.1. We shall see that it does not work for this kind
of problem. In this formulation there is a single trajectory, which begins at
$(q,p)$ and evolves with $T$. Its endpoint is given by $q_T=q-|v|T$ until it
hits the wall, which happens for time $T_{\rm r}=q/|v|$.  After this time
$q_T$ becomes $|v|T-q$. Formula (\ref{ba18}) yields for a given $T$ a single
gaussian wavepacket whose center follows this trajectory. The packet is
incoming for $T<T_{\rm r}$ and outgoing for $T>T_{\rm r}$. Incoming and
outgoing are never present at the same time and therefore there are no
interference effects. In a calculation for a soft wall, one would see the full
amplitude of the gaussian packet on the back side of the wall, with no damping
due to barrier penetration, and again no interference effects.

On the other hand, the other two approximations with real trajectories,
$(x_f,q)$ and $(x_f,p)$, do work fine for the hard wall and give the exact
answer. Each of them contains a direct and a reflected trajectory at all
times, hence a sum of two wavepackets with interference.

\section{The inverted gaussian potential}

The hard wall is probably the simplest example after the free particle.
However, except for $\psi_{qp}$, the calculation involves two trajectories.
Our aim here is to consider a potential where a single classical trajectory
contributes to the semiclassical formulae, to avoid dealing with interference
and focus only on the differences arising from the complex or real character
of a trajectory. We choose an inverted gaussian potential with hamiltonian
\begin{equation}
H(X,P) = \frac{P^2}{2} - e^{-X^2} \;.
\end{equation}
The system has a single parameter, which is the width of the wavepacket. We
set $\hbar=1$ and we place the packet initially at $q=-5$, $p=1/2$, hence
central energy 1/8. There is a single bound level with energy -0.48, which
excludes two simple limits: that in which the potential is just a
perturbation, and that in which the problem is highly semiclassical.  Here
both the real and the complex trajectories need to be calculated numerically,
as well as the exact packet propagation of course. The numerical methods for
the trajectories will be discussed in the next section.

Fig. 1 shows the wave packet at $T=7$ for the different approximations,
compared with the exact result, for two values of the width, $b=0.5$ and
$b=1.0\;.$ The packet spreads very fast and rapidly becomes highly
non-gaussian. It is remarkable how well this non-gaussian character is
reproduced by some of the approximations, though not all. We shall have more
to say in section 7 about comparing the various approximations.  The top two
graphs are the probability density versus $x_f$. The bottom two are the phase
of the wave function. Both the modulus and the phase are very well reproduced
by $\psi_{CT}\;,$ the approximation via complex trajectories, and
$\psi_{{x\!_f}q}\;,$ one of the real trajectories approximations,
$\psi_{{x\!_f}q}$ being just as good as $\psi_{CT}$ for this. On the other
hand $\psi_{qp}$, which is a pure gaussian, does not do well at all.  The
small zig-zags in the approximate calculations of the phase are due to
numerical imprecisions.  Here, as well as in the next section, we do not show
$\psi_{{x\!_f}p}\,,$ because it displays basically the same features as its
counterpart $\psi_{{x\!_f}q}\;.$ Calculations with other values of $b$ show
these results to be robust, in the sense that $\psi_{{x\!_f}q}$ is always very
similar to $\psi_{CT}$ and that both agree well with the exact propagation.

\section{The quartic oscillator}

In this section we apply our semiclassical formulae to the case of a totally
binding potential. We choose the quartic oscillator because it is probably the
simplest system after the harmonic oscillator (for which all semiclassical
formulae of sections 2 and 3 are exact). The hamiltonian is
\begin{equation}
    H = \frac{P^2}{2} + A X^2 + B X^4
    \label{s41}
\end{equation}
and the parameters are set to $A=0.5$, $B=0.1$, $\hbar=1\;.$ For these values
the ground state energy is $E_0\approx 0.559$ and the first two excited states
have $E_1\approx 1.770$ and $E_2\approx 3.319$. For the wavepacket we choose
$q=0$, $p=-2.0$, and $b=1.0$. This gives $E=H(q,p)=2.0$ for the energy of the
central trajectory, $\tau \approx 4.7$ for its period, and $X_{\rm{turn}}
\approx \pm 1.6$ for its turning points.  Fig.2 shows a sequence of four
snapshots in the exact time evolution of the packet.  The energy is low, the
wavelength is large, and the interference effects are important.

As in the previous example, both the real and the complex trajectories have to
be computed numerically.  The real trajectories from $q$ to $x_f$ can be
calculated with a simple `shooting method'. Since we need the propagated wave
packet at all values of $x_f$, we simply integrate the equations of motion
starting from $X(0)=q$ (fixed) and several values of the initial momentum
$P(0) \equiv p_i$. For each $p_i$ we obtain a final coordinate $X(T)\equiv x_f$
at which we evaluate the wave function.

The calculation of the complex trajectories is more involved and we shall
describe it in some detail. We follow the procedure introduced by Klauder
\cite{Kla87} (see also \cite{Hel87,adachi,Rub95} and \cite{Rib04} for a
different approach), which consists in propagating trajectories starting from
\begin{equation}
    X(0) = q + w \qquad \qquad P(0) = p + i \frac{c}{b} w
    \label{s42}
\end{equation}
where $w\equiv\alpha + i\beta$ is a complex number. The first of conditions
(\ref{bcf}) is automatically satisfied for all $w$. The search method consists
in propagating trajectories for all possible $w$'s and picking those
satisfying the second condition (\ref{bcf}), $X(T)=x_f\;.$

As discussed in \cite{Rub95}, fixing $q$, $p$ and $T$ and integrating
Hamilton's equations with initial conditions (\ref{s42}) gives, for each $w$,
a final coordinate $X(T)\equiv X_T$, usually complex. This can be viewed as a
map $X_T = X_T(w)$. The values of $w$ we need are given by the inverse map of
the real $X$ axis. If the map is analytic at $w$, then it is conformal. This
implies, among other things, that the map is one-to-one in the neighborhood of
$w$. If, on the other hand, $w$ is a critical point of the map, a richer
structure develops.  If, for instance, $\partial X_T/\partial w = 0$, but the
second derivative is non-zero, it can be shown that the map becomes two-to-one
in the vicinity of $w$, meaning that two different trajectories (corresponding
to two distinct initial conditions) satisfy the same boundary conditions. This
is the basic mechanism that generates multiple trajectories in systems with
one degree of freedom. For short times the critical points of the map
generically lie very far from the origin $w=0$, corresponding to complex
trajectories whose actions have large imaginary parts. As time increases, more
and more of these points approach the origin, giving rise to significant
contributions to the propagator \cite{Rub95}. The singularities in the $X_T
(w)$ map produced by the critical points are called Phase Space Caustics
(PSC). At these points the square-root in the pre-factor of Eq.(\ref{ba14})
goes to zero and the semiclassical approximation fails (it actually fails in a
finite neighborhood of the PSC's).

For fixed $q\,, p\,,\,{\rm and} \, T$, the complex trajectories
form continuous families parameterized by $x_f\,.$ Among the many
families that might contribute to the semiclassical evolution of
the wave-packet, one is special. This is the family that contains
the real trajectory that starts at $(q,p)$, and we call it the
{\it main family}. If $(q_T,p_T)$ is the end point of this real
trajectory, then for $x_f=q_T$ the trajectory in the main family
is real. In terms of the map, the point $w=0$ is mapped into
$X_T=q_T$. In some special situations, the main family alone may
provide sufficient information for the semiclassical calculation
\cite{Hel87,Hell02,Hell03}, but that is not always the case
\cite{Shu95,Shu96,Rib04}.

Fig.3(a) shows the map $X_T=X_T(w)$ for $T=6.5$. The lines correspond to
constant values of ${\rm Re}(X)$ and ${\rm Im}(X)$. The conformal property of
the map guarantees that these lines cross at right angle (provided the same
scale is used for $\alpha$ and $\beta$). The circle indicates the location of a
critical point, corresponding to a singularity in the regular pattern of the
lines. The thick lines correspond to families of trajectories for
which ${\rm Im}(X(T)) = 0\,.$ The main family can be identified by its
containing the point $w=0\;.$ A more careful look at this figure reveals the
existence of several other singularities in the otherwise regular pattern of
crossing lines. Each of them is a critical point of the map. The trajectories
in their neighborhood, however, do not contribute significantly to the
propagated wave packet, and we shall not take them into account.

Let
\begin{align}
   \label{expf}
     F \equiv S(x_f,x_0;T) + p(x_0-q/2) +
      i\hbar\frac{(x_0-q)^2}{2 b^2} \;
\end{align}
be the exponent in Eq.(\ref{ba14}). For real trajectories ${\rm
Im}(F)$ is $\ge 0$. Fig.3(b) shows a greyscale topographic map of
${\rm Im}(F)$ for the trajectories calculated in Fig.3(a) and in
the same $(\alpha,\beta)$ plane. The continuous thick line is the
main family seen on Fig.3(a). The line for the other family, which
we call {\it secondary}, is part thick and part thin. The reason
for this distinction is the following. The imaginary part of $F$
for trajectories in the main family can be seen to be always
positive. It starts with ${\rm Im}(F)=+\infty$ for $x_f =+\infty$,
decreases to zero at $x_f=q_T$ and grows to infinity again as $x_f
\rightarrow -\infty$. For the trajectories in the secondary
family, however, ${\rm Im}(F)$ starts at $+\infty$ for $x_f =
+\infty$, and decreases steadily to ${\rm Im}(F)=-\infty$ at $x_f
= -\infty$.  This means that this family cannot be included in the
semiclassical calculation for all values of $x_f$. There must be a
cutting point after which the family cannot be considered,
otherwise it would cause the wave function to diverge. This
phenomenon has become known as the {\it problem of
non-contributing trajectories}, and has been studied by Adachi
\cite{adachi}, Berry \cite{Ber89}, Klauder \cite{Rub95} and others
\cite{Shu95,Shu96,Tan98,Rib04}. The part of the secondary family
shown with the thin line has not been included in our
calculations.  We note that, in \cite{Hel87}, Huber and Heller do
a similar calculation for a wave packet moving in the Morse
potential.

To understand qualitatively the role of secondary families in the
semiclassical limit, we consider $\hbar$ to be really very small.  In this
case the $(x_f,q)$ and $(x_f,p)$ real trajectory approximations of section 3
become exact, and only the main family (actually only a small neighborhood of
the real trajectory) contributes significantly.  Demanding the approximation
to be uniformly valid as $\hbar\rightarrow 0$, and using the subscript $m$ for
the main family and $s$ for a secondary one, we must apply the following
rules: (a) Trajectories with ${\rm Im}(F_s)<0$ should be removed \cite{Rub95}.
This avoids the divergence of $\exp{\{{\rm Im}(F_s)/\hbar\}}$ in the limit
$\hbar \rightarrow 0$; (b) Trajectories with ${\rm Im}(F_s)(x_f)>0$ but ${\rm
Im}(F_s(x_f))<{\rm Im}(F_m(x_f))$ should also be removed to guarantee that, as
$\hbar \rightarrow 0$, the main family always gives the dominant contribution;
and (c) The discontinuity introduced by the sudden removal of a secondary
contribution should be minimized.  This criterion ultimately determines the
choice of the cutoff point, which is located on the so-called Stokes line
\cite{Rub95}.

Fig.4 shows the comparison between the exact calculation (thick
line) and the different approximations of sections 2 and 3 for
$T=2.5$ and $T=4.5\;.$ The most interesting feature here is that the
real trajectory formula $\psi_{{x\!_f}q}$ becomes discontinuous. This
can be understood with the help of panels (b) and (d) which show
the initial momentum $p_i$ of the contributing trajectories as a
function of $x_f$ (for fixed $q$ and $T$). We see that several
branches appear, but it turns out that only the one shown with a
thick line contributes significantly. Since this main branch covers only a
finite range of $x_f$, the wave function drops to zero suddenly. In fact,
the wave function diverges at the ends of the branches. In our
calculations we have cut the branch a little before its end
points.

Fig.5(a) shows the real part of $\psi_{CT}(x_f,z;T)$ for $T=6.5$
calculated with the separate contribution of each of the families
shown in Figs.3. The abrupt cutoff of the secondary family is
clear in this figure. Fig.5(b) shows again the separate
contributions of the same families to $|\psi_{CT}(x_f,z;T)|^2$ (dashed
and solid lines) and their combined contribution (thick solid
line). Notice the interference between the two families producing
the oscillation in the probability density.

Finally Fig.6 shows $|\psi_{CT}(x_f,z;T)|^2$, the calculation with
complex trajectories. Each snapshot shows the exact result (thin
solid line), the thawed gaussian approximation $\psi_{qp}$ (dashed
line) and the complex trajectories approximation (thick line). We
do not show $\psi_{{x\!_f}q}$ in these plots because the
approximation is not good. The improvement obtained with the
complex trajectories is clear. For $T=6.5$ and $T=8.5$, however,
we can see a spurious peak close to the turning points. This might
seem to result from cutting off the secondary family at the wrong
point, as suggested in \cite{Hel88}. A careful analysis shows that
it is not the case: cutting off the secondary family at a smaller
value of $x_f$ would produce a sudden dip in the probability
density followed by another sudden increase, as can be seen from
Fig.5(b). The spurious peaks are due to the phase space caustic
that shows up close to the turning points. For the present
situation, which is not quite in the semiclassical domain, the
caustics have a strong effect on the semiclassical propagations
and further corrections are necessary to improve the
approximation. A uniform semiclassical approximation involving
Airy functions can be derived by considering cubic terms in the
saddle point approximation performed in the integral (\ref{ba7}).
The derivation of this improved formula will be published
elsewhere.

\section{Summary and discussion}

We begin by summarizing the results. We have derived and written down in this
paper (sections 2 and 3) four semiclassical expressions for the propagation of
an initially gaussian wavepacket, $\psi_{\rm CT} ,\psi_{qp}\,,\psi_{{x\!_f}q}
\,,\psi_{{x\!_f}p}\;.$ All four give exact results for the free particle and
the harmonic oscillator. All except $\psi_{qp}$ also give the exact result for
the scattering by a hard wall. But $\psi_{qp}$ is very bad for the hard wall:
it produces no interference. We tested the formulae further on two examples of
smooth potentials. For the attractive gaussian potential, the exact packet
after some time is highly non-gaussian. Both $\psi_{\rm CT}$ and
$\psi_{{x\!_f}q}$ involve a single trajectory for each $x_f$, and both give
very good results for the modulus as well as the phase of the wave packet. The
same is true of $\psi_{{x\!_f}p}\;.$ But $\psi_{qp}\,,$ which is a pure
gaussian, gives a very bad fit.

The other smooth potential is a quartic oscillator. Now there may be
contributions from multiple families of trajectories and the results are quite
different. We found in this case that the complex trajectories formula
$\psi_{\rm CT}$ is superior to both the real trajectories approximation
$\psi_{{x\!_f}q}$ and the single trajectory approximation $\psi_{qp}$. One
reason is that the real trajectories contributing to $\psi_{{x\!_f}q}$ have
finite branches, which causes a drastic discontinuity in the approximate wave
function. The complex trajectories, on the other hand, form continuous
branches that prolong into the complex domain. The projection of these complex
families onto the real plane show that they follow closely the main branch of
the real trajectories, but they continue to exist after the latter end.

Thus the paper's main finding is that the complex trajectory approximation,
$\psi_{\rm CT}$ of eq. (\ref{ba14}), gives very good results, at least for all
the cases that we looked at. And one may well wonder how it could be that a
relatively simple approximation, based on a single trajectory, works so
well. The short answer is that, actually, the approximation is not that simple
and it is not based on a single trajectory. But we shall go into more details.

Many other approaches have been proposed for the propagation of an initial
wave function, some of them especially concerned with the propagation in
systems with many degrees of freedom. Among them, Initial Value
Representations or IVR's \cite{Mill74,Herm84,Hell91,Kay94a,Pol03,Tho04} have
become especially popular . They are integral expressions over the phase space
of initial conditions, in a spirit which tends to rejoin that of path
integrals.  The multiple integration is usually handled by Monte Carlo
techniques \cite{Kay94b,Zha03}. The initial impetus for these approaches
\cite{Mill70} was the root-search problem, i.e. the fact that the Van Vleck
expression (\ref{ba5}) requires one to find trajectories determined by mixed
initial-final boundary conditions (starting at $x_i$ and ending at $x_f$).
This may lead to a difficult search in complicated situations, and in addition
the root-search problem usually has several solutions, a fact which
complicates further the time-development of the wave function. In the IVR's,
on the other hand, the trajectories are determined solely by initial boundary
conditions, as the name implies, and for each set of initial conditions there
is only one trajectory, the unique solution of Hamilton's equations for these
conditions. The work of getting the correct propagated wave function is done
entirely by the integration, as is the case for path integral expressions.

The present work is based on the Van Vleck expression and
it is affected by root-search, except for the thawed gaussian formula
$\psi_{qp}$ of section 3.1. Root-search can truly be a burden, especially for
multi-dimensional wave functions. More important, however, is the fact that
characterizing the present approach as a one-trajectory or a few-trajectories
approximation is highly misleading. We have used such characterization several
times above, for instance in connection with the hard wall, or in the first
few lines of section 5. But the truth is that there is a different classical
trajectory for every final $x_f$, and therefore the number of trajectories is
really infinite. The work of getting the correct propagated wave function is
done by them.

The one case where we truly have a one-trajectory expression is $\psi_{qp}\;.$
There, the same trajectory is used for all values of $x_f$. This case is truly
an IVR, though a very simplified one, not involving any integration over the
initial parameters. The only initial parameters used are those of the center
of the packet. And it is also the case for which, in our explorations, the
agreement with the exact result is worst, showing that a single classical
trajectory is just not enough. Somehow, a variety of classical trajectories
must enter into the formalism.  In other words, we see the path integral idea
reappear, Feynman's original view of the connection between quantum and
classical mechanics.

Among the various IVR's, there is one that has been especially successful and
has had many applications, the Herman-Kluk approximation
\cite{Herm84,Kay94a,Gar00,Wan01,Noi03}.  It would be particularly interesting
to have a direct application of the HK formula to the hard wall, since this is
the simplest non-obvious case for which our complex trajectories formula is
exact. We are not aware of the existence of such an application of HK, and it
seems to us that the calculation will not be totally straightforward.

Perhaps it is not superfluous to repeat once again that the main new result of
this paper is that semiclassical approximations using complex trajectories,
especially non-IVR approximations, are capable of being surprisingly good. All
our semiclassical expressions generalize easily to many dimensions. The
detailed presentation of these generalizations will be published separately,
together with numerical applications. We note that two-dimensional
semiclassical calculations with complex trajectories have already been
considered in the context of coherent states.  In \cite{Hell02,Rib04} the
quantity evaluated was the return probability amplitude, also known as the
auto-correlation function.  In \cite{Shu95,Shu96,Oni01} the authors, who seem
to have been unaware of references \cite{Hel87,Hel88}, calculate the
propagated wave-function in the momentum or coordinate representation.  But
they do it for quantum maps, which is rather different from what we have done
here since, in their case, there are no continuous classical trajectories, no
Hamilton equations, and no Schroedinger equation.  In the present paper we
provided formulae for the calculation of the propagated wave-function for
continuous systems in the coordinate representation. The basic equation for
such calculations is the complex trajectory formula (\ref{ba14}) first
derived by Heller and coworkers in \cite{Hel88}.  We showed that one can
approximate the complex trajectories by real ones, and that this can be done in
various ways (we mention three of them), with different degrees of difficulty
and of accuracy. The various formulae so generated will undoubtedly have
different domains of applicability. And the complex formula, if it is
practical, will always remain the better one.

As our final topic, we shall present an alternative derivation of a formula
very similar to (\ref{ba14}), which possesses the same three approximations in
terms of real trajectories. The quantity we want to calculate is $\langle
x_f|K(T)|z\rangle$. To obtain the approximation (\ref{ba3a}), we introduced
between $K(T)$ and $|z\rangle$ the resolution of unity in terms of eigenstates
of $x$, and then we replaced the exact $x$-propagator by its Van Vleck
approximation. Instead of this, we shall introduce between $\langle x_f|$ and
$K(T)$ the resolution of unity in terms of coherent states, thus
\begin{align}
  \label{conc1}
      \langle x_f|K(T)|z\rangle
   =  \int\langle x_f|z'\rangle
      \frac{{\rm d}^2z'}{\pi}\langle z'|K(T)|z\rangle
\end{align}
and then we shall replace the exact coherent state propagator by its
semiclassical approximation. The latter was worked out in detail in
\cite{Bar01}. Performing the integrals over $q'$ and $p'$ by the stationary
exponent approximation, we arrive at an expression which is very similar, but
not identical, to (\ref{ba14}). There are two important differences. One is
that, in this case, the dynamics are governed by the smoothed hamiltonian
${\cal H}(q,p) = \langle z|H|z\rangle$ instead of the classical hamiltonian.
The other is that the exponent acquires an extra term given by
\cite{Bar01}
\begin{align}
 \label{conc2}
   \frac{i}{\hbar} {\cal I} \equiv
   \frac{i}{\hbar} \int_{0}^{T} {\rm d}t \left(
   \frac{b^2}{4} \frac{\partial^2\cal H}{\partial q^2} +
   \frac{c^2}{4} \frac{\partial^2\cal H}{\partial p^2} \right)\; .
\end{align}
This term plays a very important role and cannot be dropped from
the approximation. Both ${\cal I}$ and the ${\cal H}$ dynamics get
carried over into all the real trajectories approximations.

Our main reason for mentioning this alternative is to stress that
semiclassical approximations are never unique and that there are
always many ways to base an approximate quantal expression on a
classical solution. Discussions of this appear in \cite{Kla85},
who see it as a consequence of the over-completeness of the
coherent-state basis, and also in \cite{Bar01}. It is only in the
limit of $\hbar$ going to zero that they all become the same. From
our alternative to (\ref{ba14}) involving the smoothed Hamiltonian
and the correction term ${\cal I}$, one can derive again three
expressions in terms of real trajectories, different from those in
section 3. Numerically, the two points of view yield roughly
equally good (or sometimes equally bad) results. And once again,
for quadratic Hamiltonians, all the approximations we have
mentioned are exact. While in section 3 the $\psi_{qp}$
approximation was the same as Heller's TGA, for the alternative
formulation it is the same as the approximation in section 4 of
\cite{Bar01}. The latter reference contains a discussion of the
differences between the two results. Both are initial value
formulae using the real central trajectory of the packet, and both
are gaussian in shape for all potentials and all times. Both are
expected to fail if the potential is not very smooth on the scale
of the size of the packet.\\

\noindent Acknowledgements

\noindent This work was partly supported by the Brazilian agencies
FAPESP and CNPq.

\newpage
\noindent REFERENCES \\

\bibliographystyle{alphamt}
\bibliography{}

\newpage

\begin{figure}
  \includegraphics[width=6cm,angle=-90]{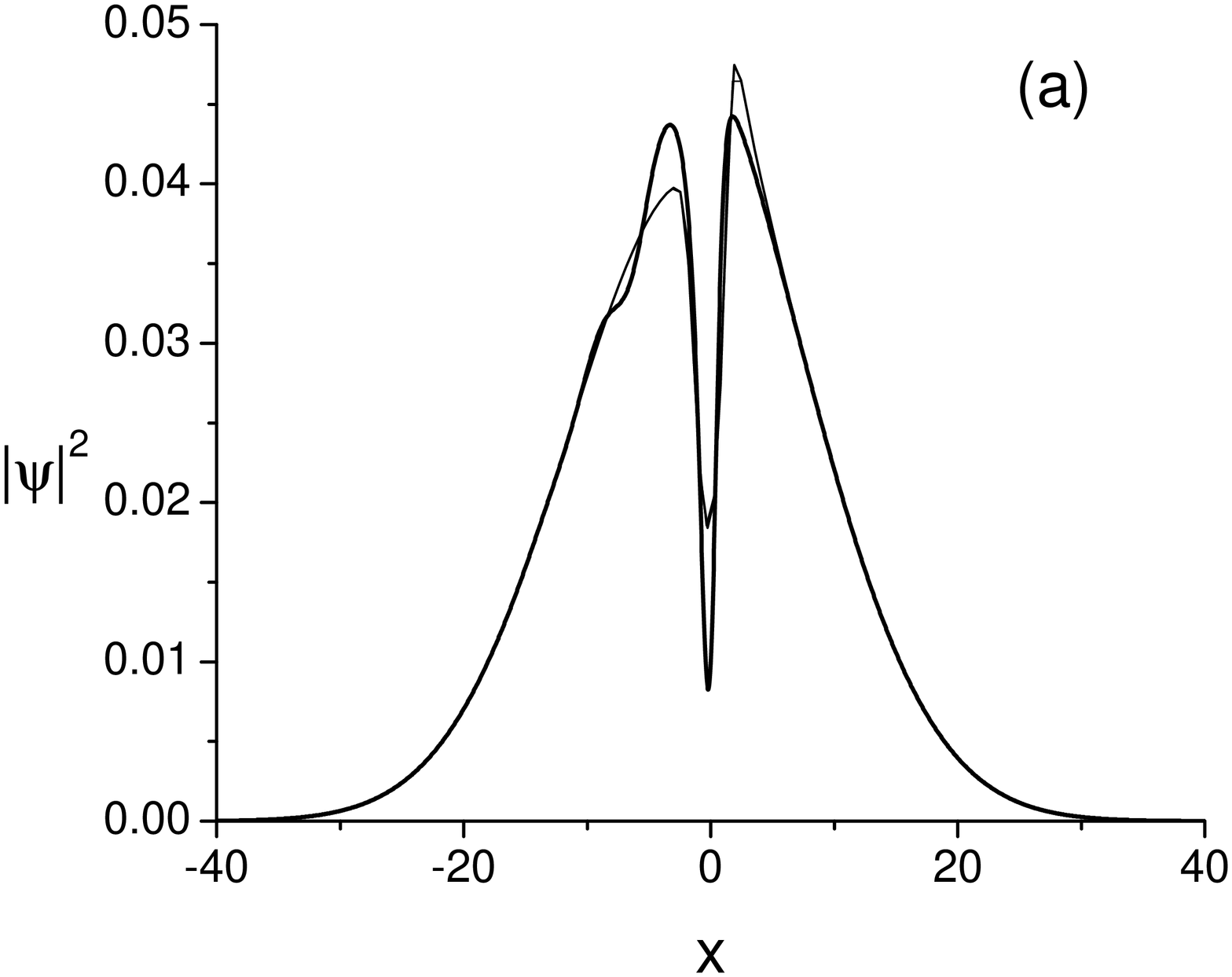}
  \includegraphics[width=6cm,angle=-90]{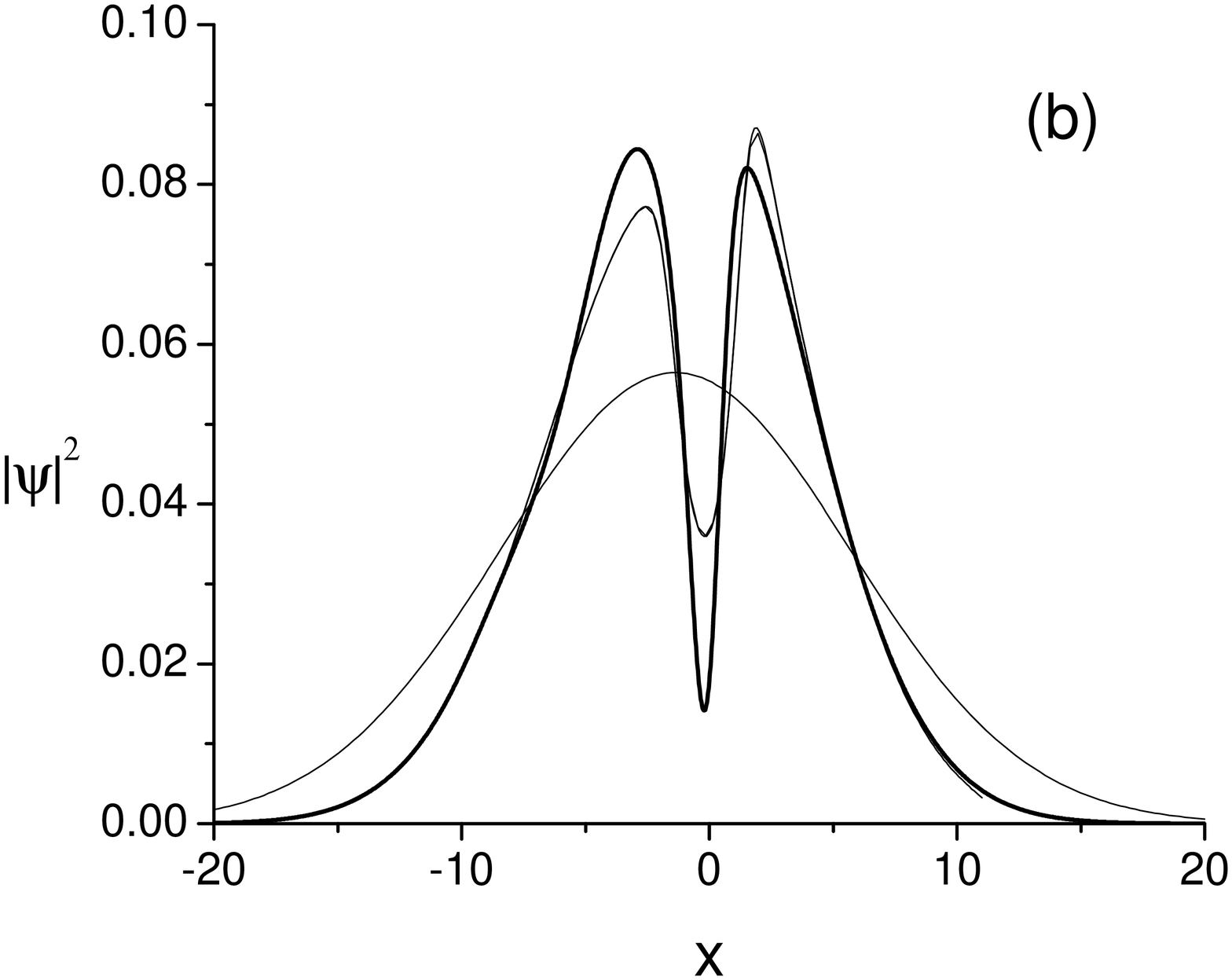}
  \includegraphics[width=5.5cm,angle=-90]{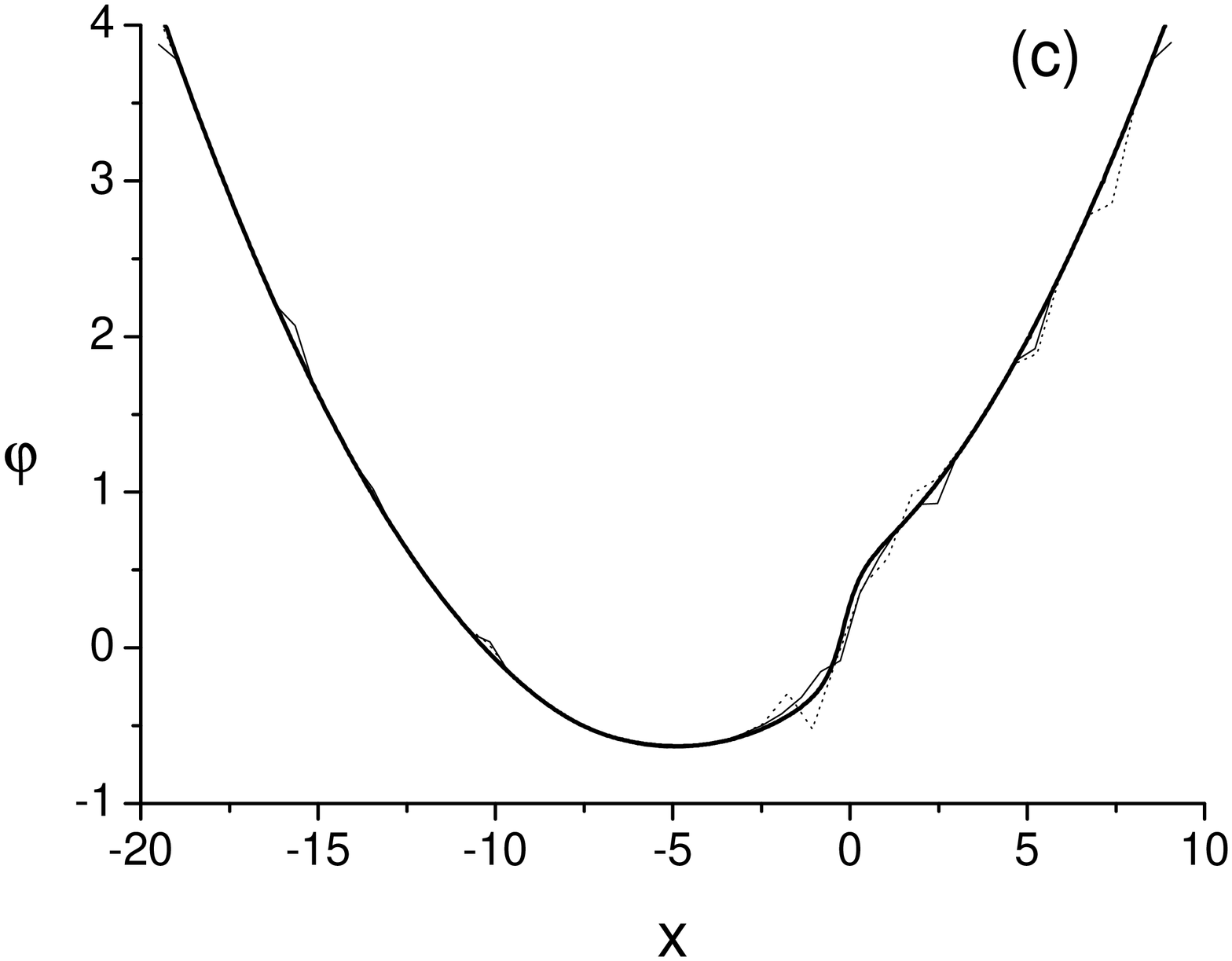}
  \includegraphics[width=5.5cm,angle=-90]{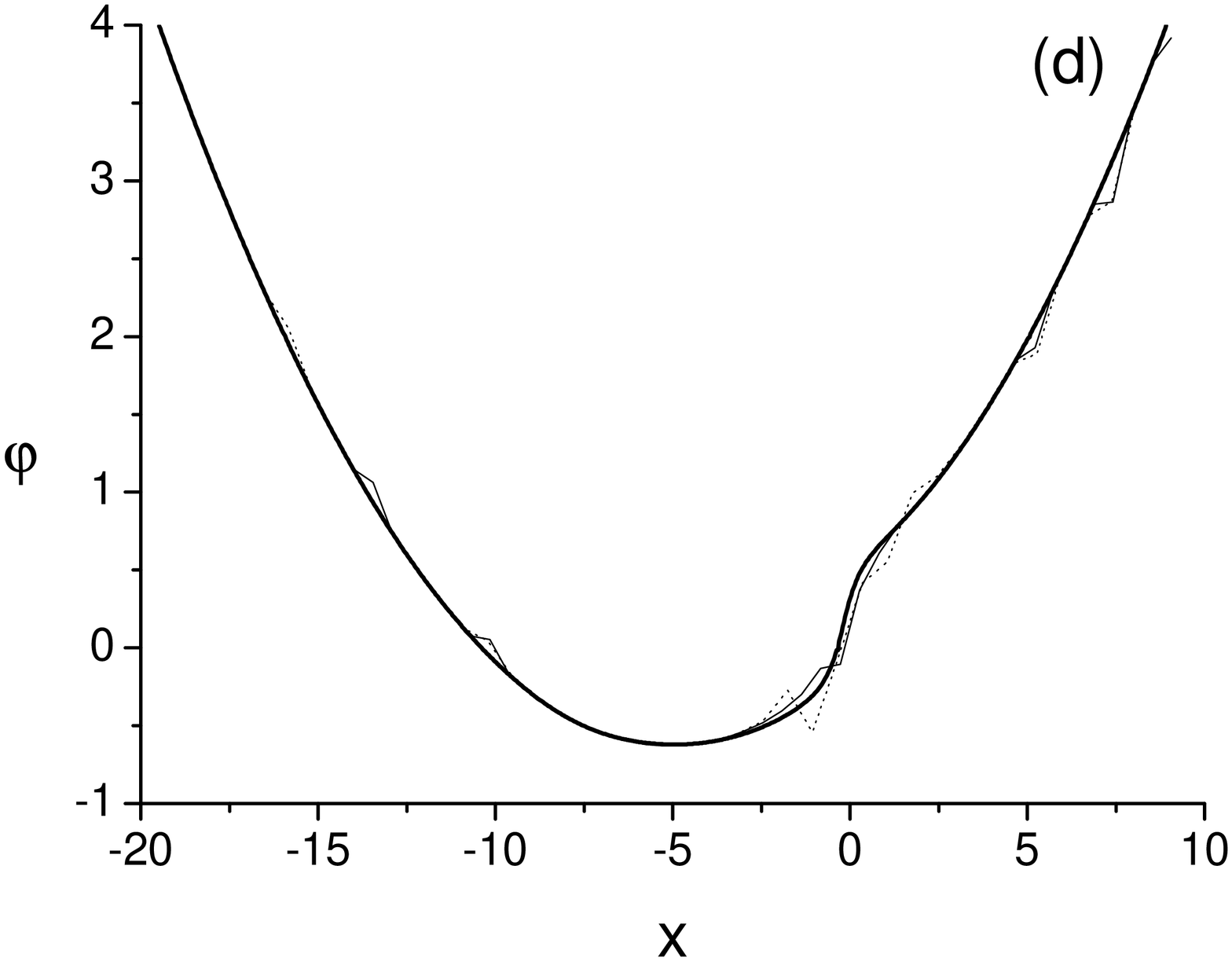}\\
 \caption{Parts (a) and (b) show the probability density in the
  inverted Gaussian potential for $p=0.5$ and $T=7.0$. In (a) $b=0.5$
  and in (b) $b=1.0$. The thick solid line shows the exact result, the
  thin solid line correspond to $\psi_{\rm CT}$ and the dotted line to
  $\psi_{x_{\!f}q}$ (which cannot be distinguished from $\psi_{\rm
  CT}$ at this scale). Part (b) also shows a comparison with $\psi_{qp}$
  (symmetric Gaussian curve). Parts (c) and (d) show the phase of the
  wave packet, in units of $\pi$, corresponding to (a) and (b) respectively.
  In this figure and those following, the $x$ variable in the
  abscissa is actually the variable $x_{\!f}$ of the text.}
  \label{fig1}
\end{figure}

\begin{figure}
  \includegraphics[width=12cm,angle=-90]{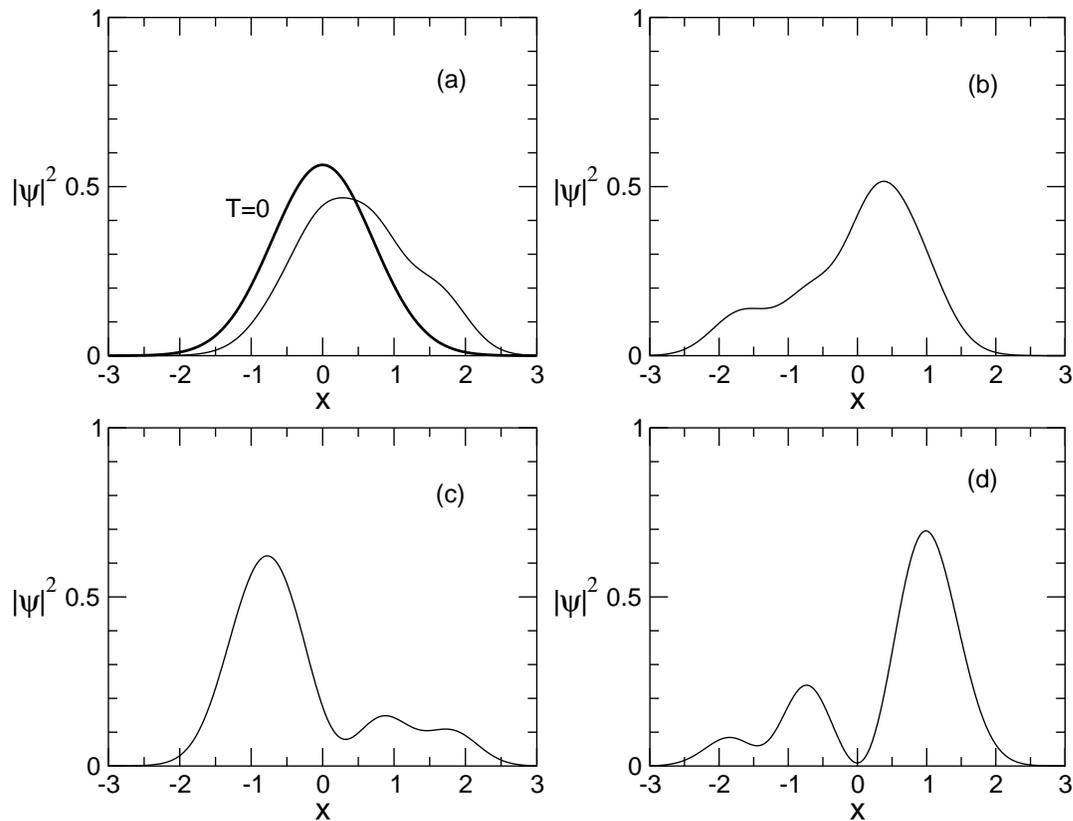}\\
  \caption{Exact quantum mechanical propagation in the quartic potential
  with $A=0.5$, $B=0.1$ and $\hbar=1$. The wavepacket
  is initially centered at $q=0$, $p=-2$ and has width $b=1$. The
  curves show the probability density at times (a) $T=0$ (thick line) and
  $T=2.5$, (b) $T=4.5$, (c) $T=6.5$ and (d) $T=8.5$. The period of the
  classical orbit of the center of the packet is $\tau \approx 4.7$. }
  \label{fig2}
\end{figure}

\clearpage

\begin{figure}
  \includegraphics[width=6.2cm,clip,angle=-90]{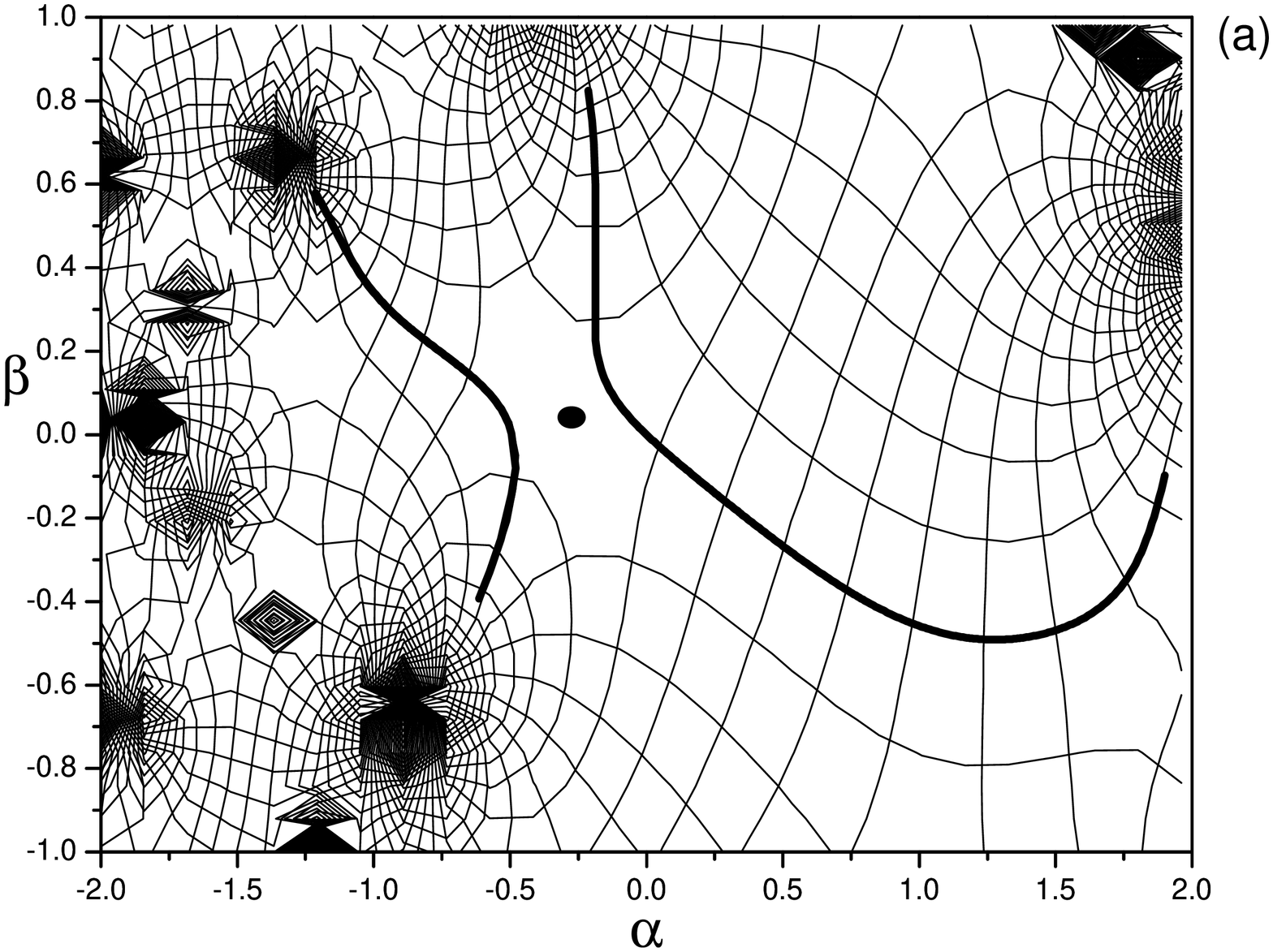}
  \includegraphics[width=7.2cm,clip,angle=-90]{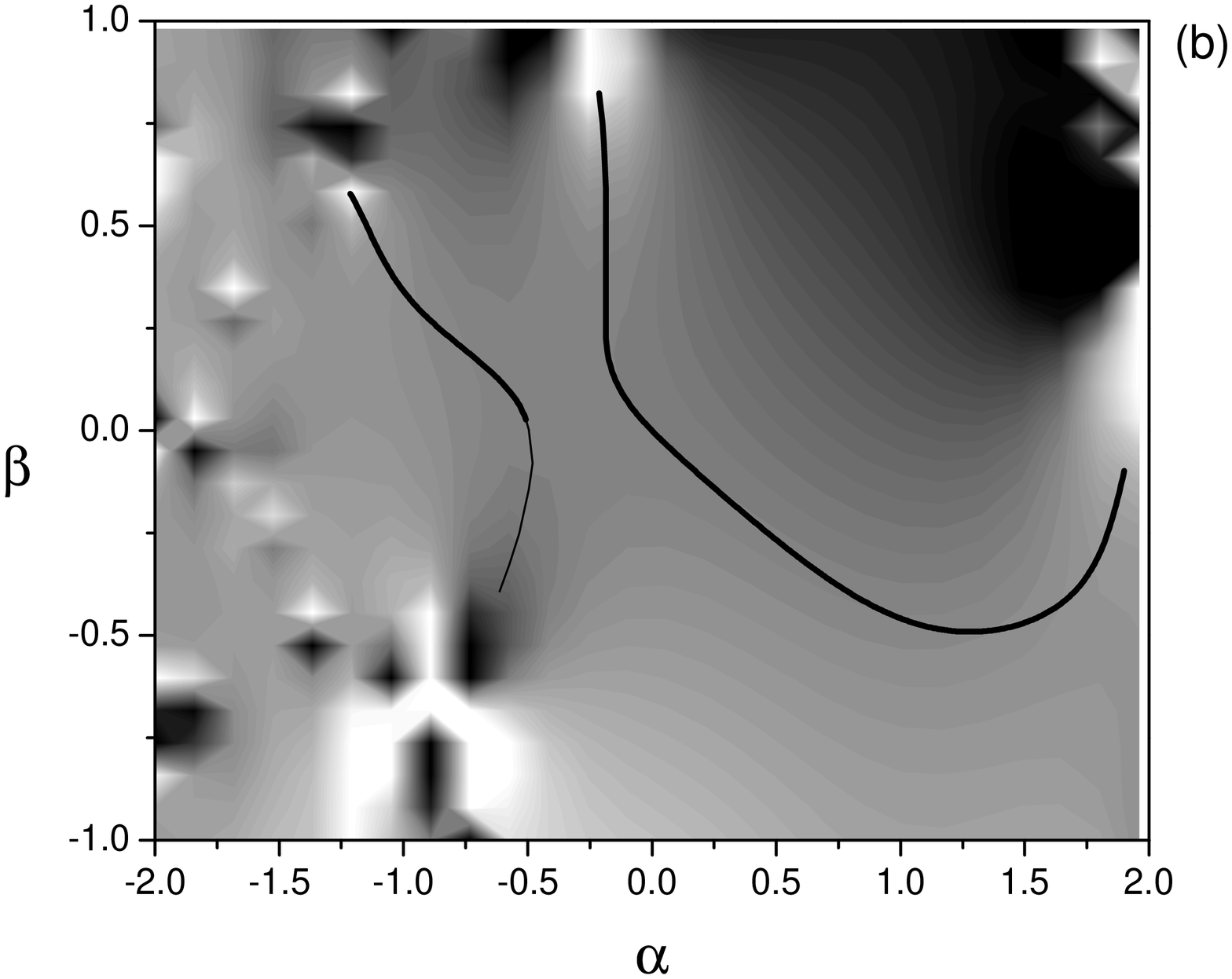}\\
  \caption{(a) Map $X_T=X_T(w)$ for $T=6.5$. The lines correspond to
  constant values of Re($X$) and Im($X$) and the circle indicates the
  singularity. The thick lines correspond to the trajectories
  satisfying Im$(X_T)=0$; (b) Gray scale topographic plot of the imaginary
  part of the exponent $F$ for all the trajectories in (a). The shades
  of gray go from $-\infty$ (black) to $+\infty$ (white). The main
  family has Im$(F) \geq 0$ whereas the secondary family has a section
  where Im$(F) < 0$ (thin line).}
  \label{fig3}
\end{figure}

\begin{figure}
  \includegraphics[width=6cm,angle=-90]{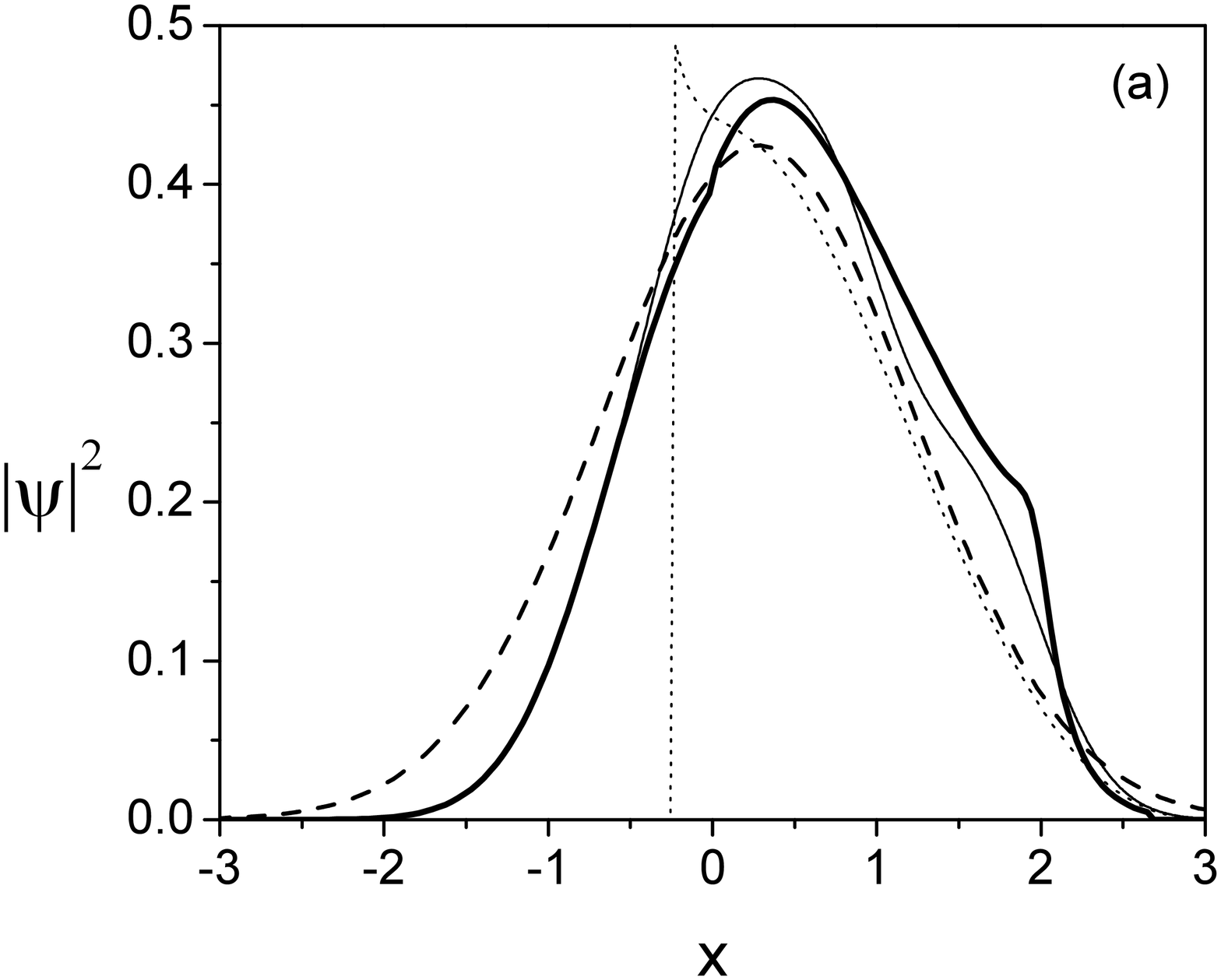}
  \includegraphics[width=6cm,angle=-90]{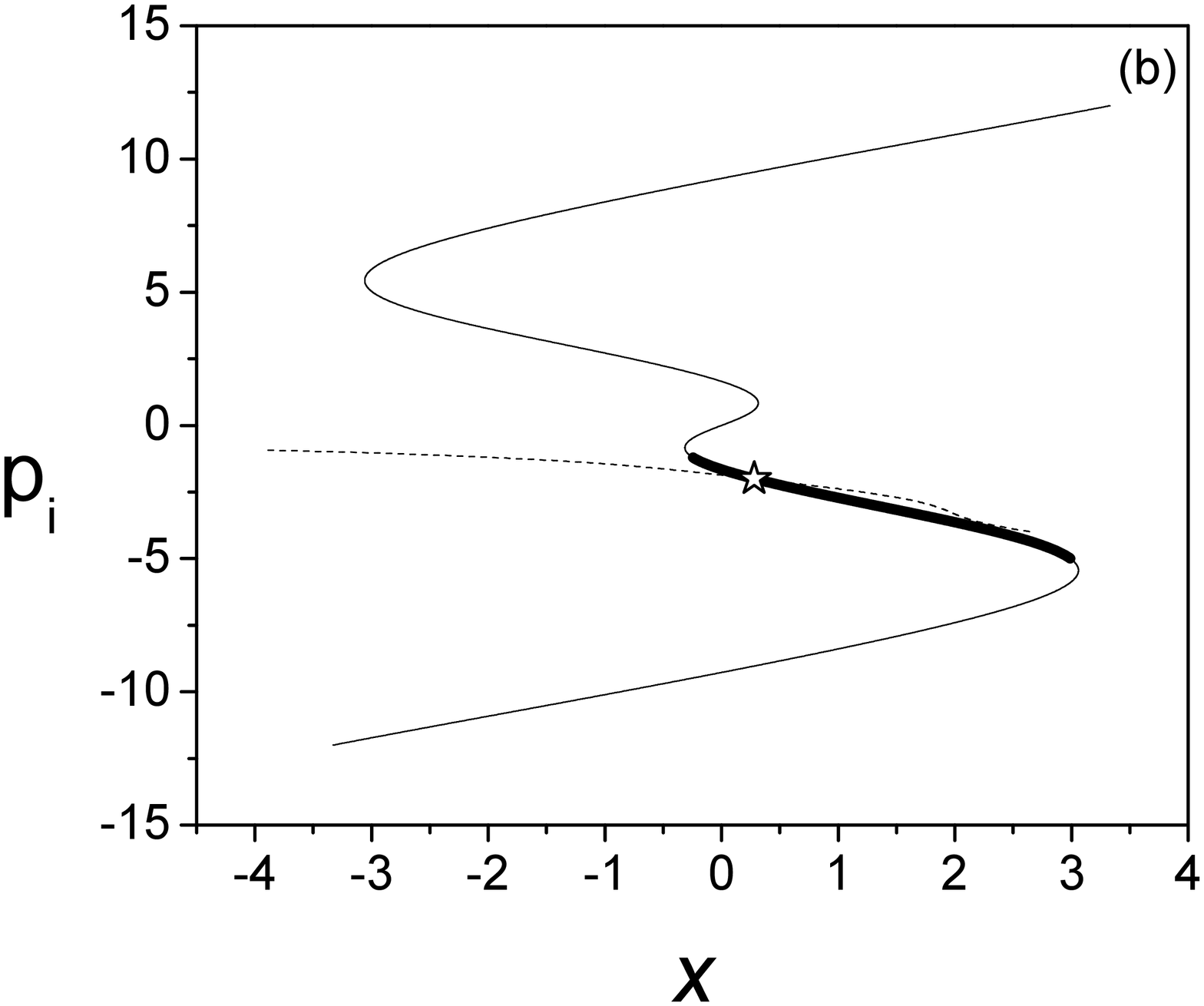}
  \includegraphics[width=6cm,angle=-90]{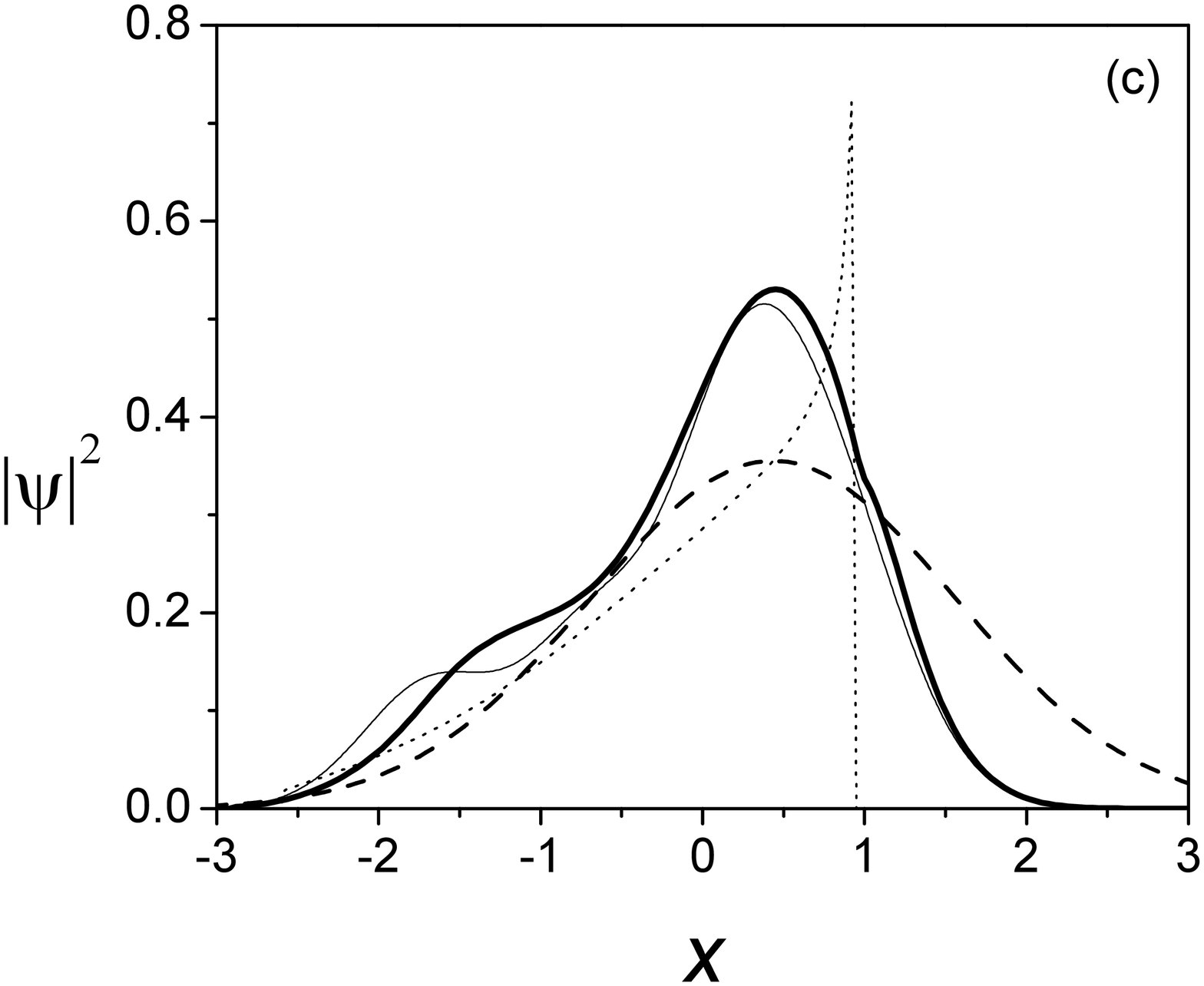}
  \includegraphics[width=6cm,angle=-90]{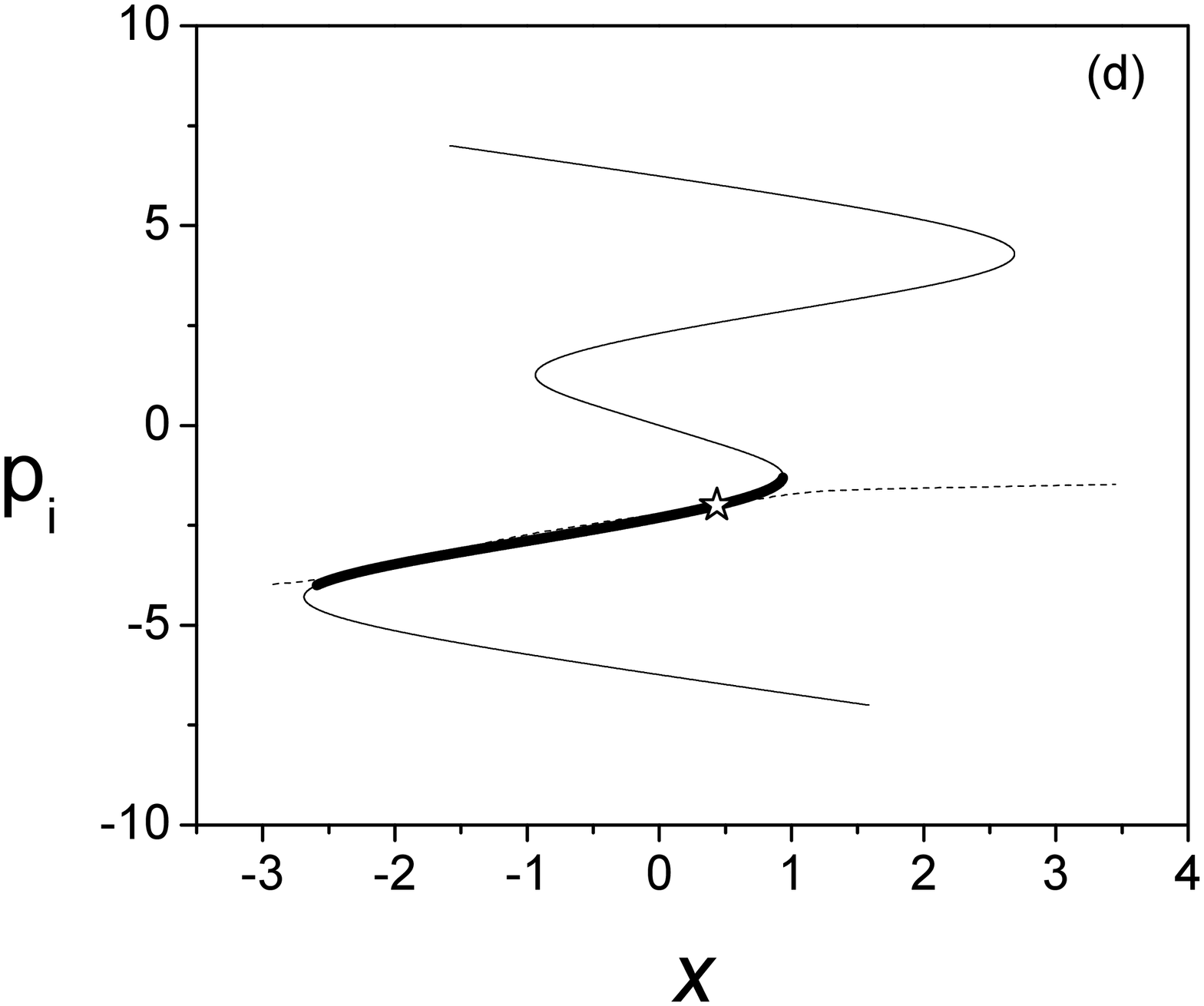}\\
  \caption{Parts (a) and (c) show the probability density in the
  quartic potential for $T=2.5$ and $T=4.5$ respectively. The thin
  solid line shows the exact result, the dotted line corresponds to
  $\psi_{x_{\!f}q}$, the dashed line to $\psi_{qp}$ and the thick
  solid line to $\psi_{\rm{CT}}$. Parts (b) and (d) show the initial
  momentum $p_i$ as a function of $x$ for the $(q,x)$ real trajectory
  for $T=2.5$ and $T=4.5$. The thick line shows the branch used in the
  calculation of $\psi_{x_{\!f}q}$ and the star represents the central
  trajectory starting from $q,p$. The dashed line corresponds to the
  projection of the main family of complex trajectories into this real
  plane.}
  \label{fig4}
\end{figure}

\begin{figure}
  \includegraphics[width=6cm,angle=-90]{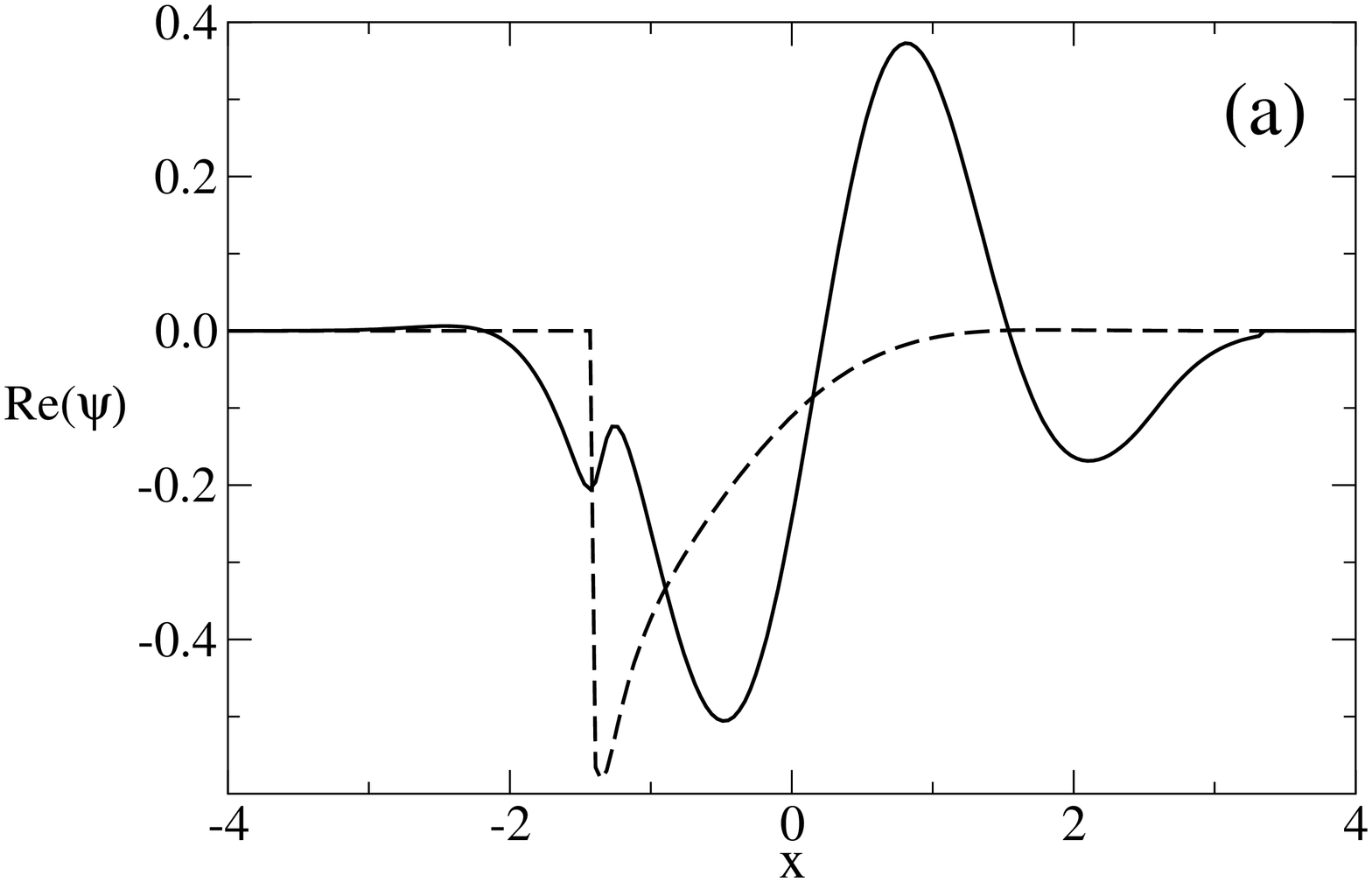}
  \includegraphics[width=6cm,angle=-90]{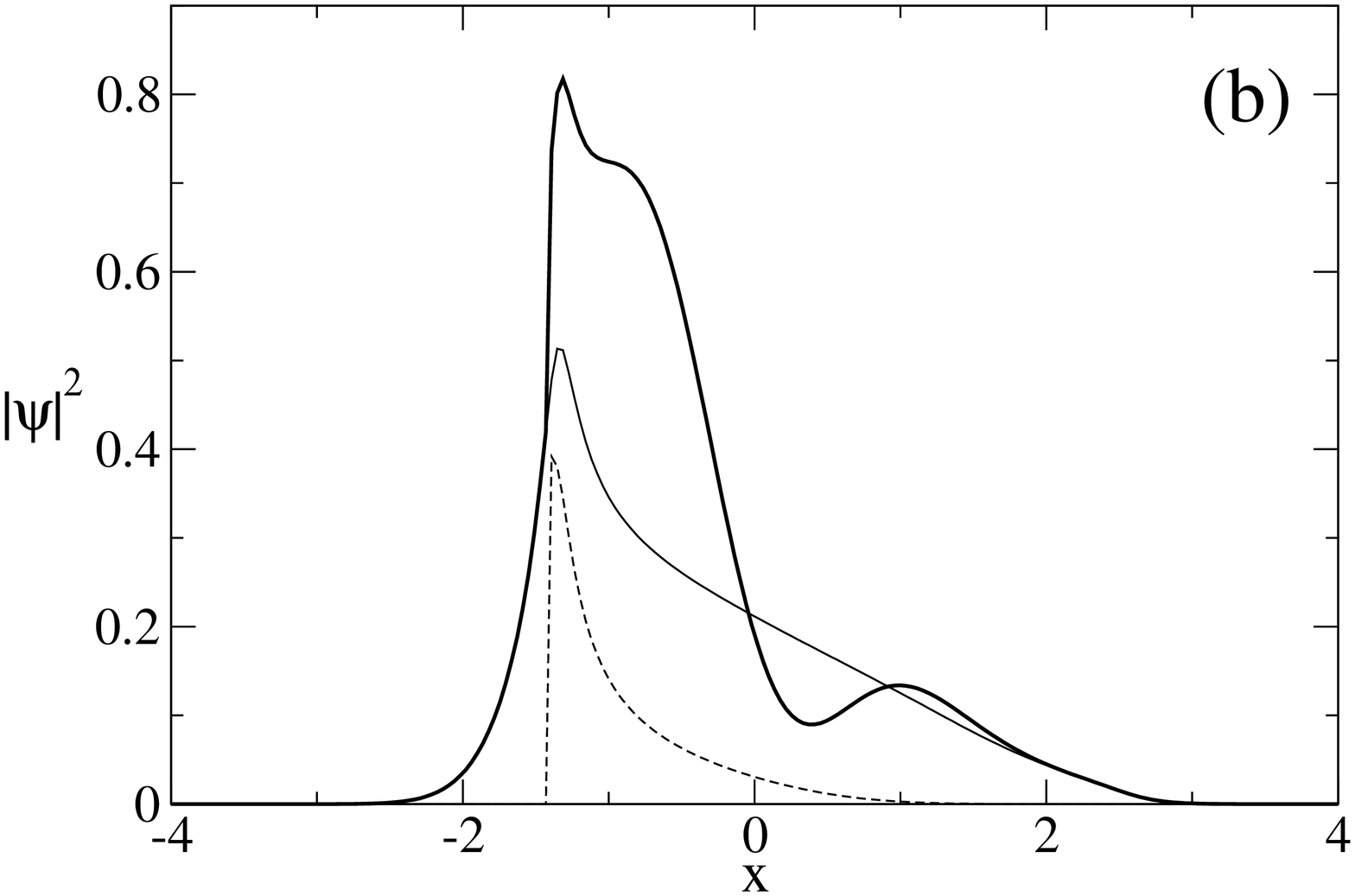}\\
  \caption{Separate contributions of the main (solid line) and
  secondary (dashed line) families for the wave packet at $T=6.5$. (a)
  Real part of $\psi$; (b)probability density $|\psi|^2$. Notice the
  abrupt cutoff of the secondary family. The thick line in (b) shows
  the total probability density, displaying the interference between
  the individual contributions.}
\label{fig5}
\end{figure}

\begin{figure}
  \includegraphics[width=12cm,angle=-90]{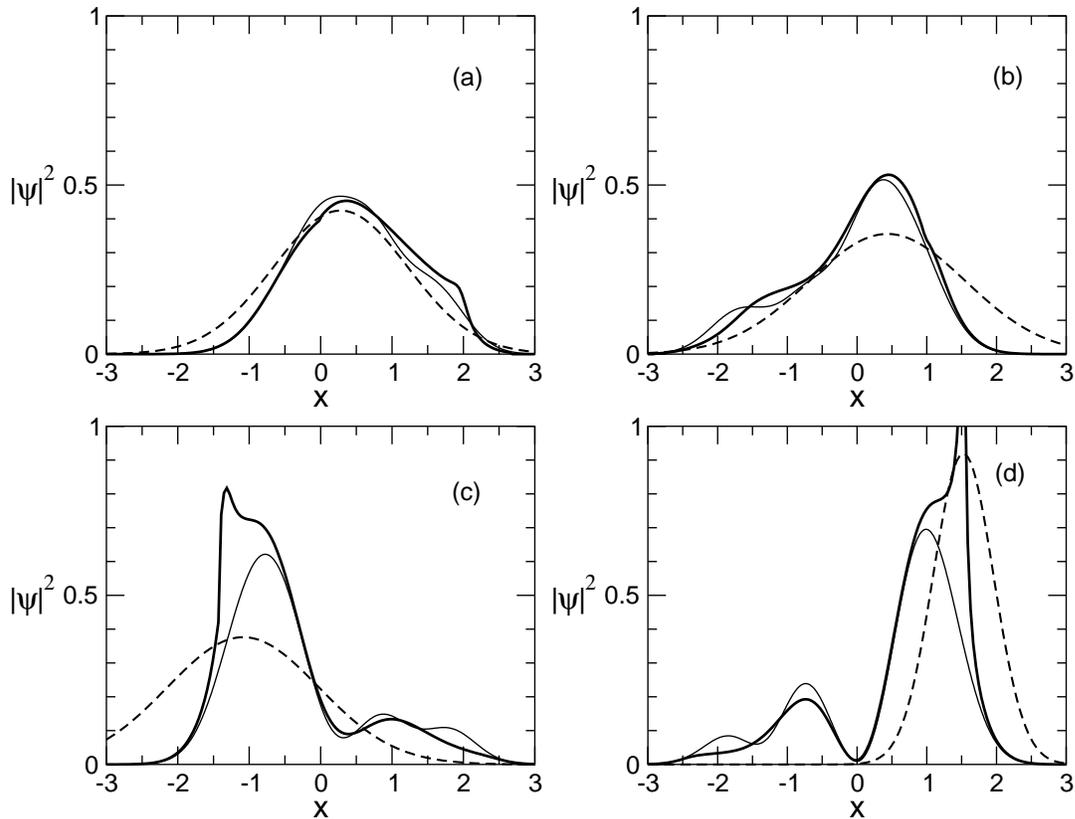}\\
  \caption{Probability density for the quartic potential for (a)
  $T=2.5$; (b) $T=4.5$; (c) $T=6.5$ and (d) $T=8.5$. The thin solid
  line shows the exact result, the dashed line corresponds to
  $\psi_{qp}$ and the thick solid line to $\psi_{\rm{CT}}$.}
  \label{fig6}
\end{figure}

\end{document}